\documentclass[amsmath,amssymb,aps,twocolumn,footinbib,superscriptaddress]{revtex4-1}

\usepackage[bookmarks=false,linkcolor=blue,urlcolor=blue,colorlinks,citecolor=blue]{hyperref}
\usepackage{graphicx}
\usepackage{bm}
\usepackage{braket}
\usepackage{color, soul}
\usepackage{amsmath,amssymb}

\begin{document}

\title{Learning phase transitions from dynamics}

\author{Evert van Nieuwenburg}
\thanks{Equal contribution}
\affiliation{Institute for Quantum Information and Matter, Caltech, Pasadena, California 91125, USA}
\author{Eyal Bairey}
\thanks{Equal contribution}
\affiliation{Physics Department, Technion, 3200003, Haifa, Israel}
\author{Gil Refael}
\affiliation{Institute for Quantum Information and Matter, Caltech, Pasadena, California 91125, USA}

\begin{abstract}
	We propose the use of recurrent neural networks for classifying phases of matter based on the dynamics of experimentally accessible observables. We demonstrate this approach by training recurrent networks on the magnetization traces of two distinct models of one-dimensional disordered and interacting spin chains. The obtained phase diagram for a well-studied model of the many-body localization transition shows excellent agreement with previously known results obtained from time-independent entanglement spectra. For a periodically-driven model featuring an inherently dynamical time-crystalline phase, the phase diagram that our network traces in a previously-unexplored regime coincides with an order parameter for its expected phases.
\end{abstract}

\maketitle

\emph{Introduction} - Machine learning is emerging as a novel tool for identifying phases of matter \cite{Carrasquilla2016,Wang2016,VanNieuwenburg2017,Broecker2017,Wetzel2017,Wetzel2017a,LiuNieuwenburg2017,Chng2016,Chng2018,Broecker2016,Schindler2017,Ohtsuki2016,Arsenault2014a,Arsenault2015,Beach2017}. At its core, this problem can be cast as a classification problem in which data obtained from physical systems are assigned a class (i.e. a phase) using machine learning methods. This approach has enabled autonomous detection of order parameters~\cite{Wang2016,Wetzel2017,Wetzel2017a}, phase transitions~\cite{Carrasquilla2016,VanNieuwenburg2017} and entire phase diagrams~\cite{Broecker2017,LiuNieuwenburg2017,Yoshioka2017,Venderley2017}. Simultaneous reserach effort at the interface between machine learning and many-body physics has focussed on the use of neural networks for efficient representations of quantum wavefunctions~\cite{Carleo2016,Schmitt2017,Cai2017,Huang2017, Deng2016, Nomura2017, Deng2017, Gao2017, Torlai2017}, drawing a parallel between deep networks and the renormalization group~\cite{Mehta2014, Koch-Janusz2017, Wang2018}. Overall, these studies exemplify the power of machine learning for extracting information from physical data without detailed physical input. In particular, it shows potential for identifying novel phases through automatic processing of large-scale data; possibly identifying features that may have been missed before.

So far, however, these methods have relied only on static properties of the underlying physical systems, such as raw state configurations sampled from Monte Carlo simulations \cite{Carrasquilla2016, Beach2017} or entanglement spectra obtained using exact diagonalization~\cite{VanNieuwenburg2017,Schindler2017,Venderley2017}. To our knowledge, the study of phase transitions from dynamics of physical observables has not been adressed.

Here, we suggest a machine learning approach to distinguish between phases based on dynamics of measurable quantities. Specifically, we introduce the use of recurrent neural networks (RNNs), designed for processing sequential data such as time-traces. This approach does not rely on thermal equilibrium, and applies very naturally to time-dependent systems. It is therefore particularly suited for the identification of dynamical as well as Floquet phases \cite{Moessner2017, Lindner2011, Titum2016, Nathan2017, Else2017, Heyl2017, vonKeyserlingk2016b, Else2016b, Po2016, Potter2016}.

We first test our method on a system with two inherently different dynamical behaviours, namely a 1D system with a many-body localization transition \cite{Basko2006, Oganesyan2007, Nandkishore2015, Abanin2017}. Machine learning methods applied on entanglement spectra of eigenstates were used to obtain a phase diagram of the same model~\cite{Schindler2017}, as well as on a slightly different model featuring two distinct MBL phases~\cite{Venderley2017}. Here, we insist on using only experimentally relevant (i.e. measureable) quantities such as the magnetization of individual spins. We find that the network succeeds at distinguishing between the ergodic and localized phases of this model, recovering phase boundaries similar to those obtained by previous methods.

We then apply our method to a periodically driven model, featuring among its three phases one which is unique to the time-dependent setting, namely a time crystal \cite{Else2016,Khemani2016,Yao2017, Wilczek2012, Choi2017, Zhang2017, vonKeyserlingk2016}. Indeed the method distinguishes between the time-crystalline, Floquet-ergodic and Floquet-MBL \cite{Ponte2015, Lazarides2015, Bordia2017} phases of this model.

In the following section, we first introduce the essentials of recurrent neural networks. We refer the reader to Ref.~\cite{Nielsen2015} for an extensive introduction to the non-recurrent feed-forward neural network. After we have introduced the network essentials, we outline the procedure we refer to as `blanking' for training the network on a set of physics data. This framework is independent of the underlying model, and serves as the main supervised learning scheme in our work. Next we turn to introducing the models and the results mentioned earlier, and conclude with a critical evaluation of the obtained results.

\begin{figure}[h!]
    \begin{center}
        \includegraphics[width=0.45\textwidth]{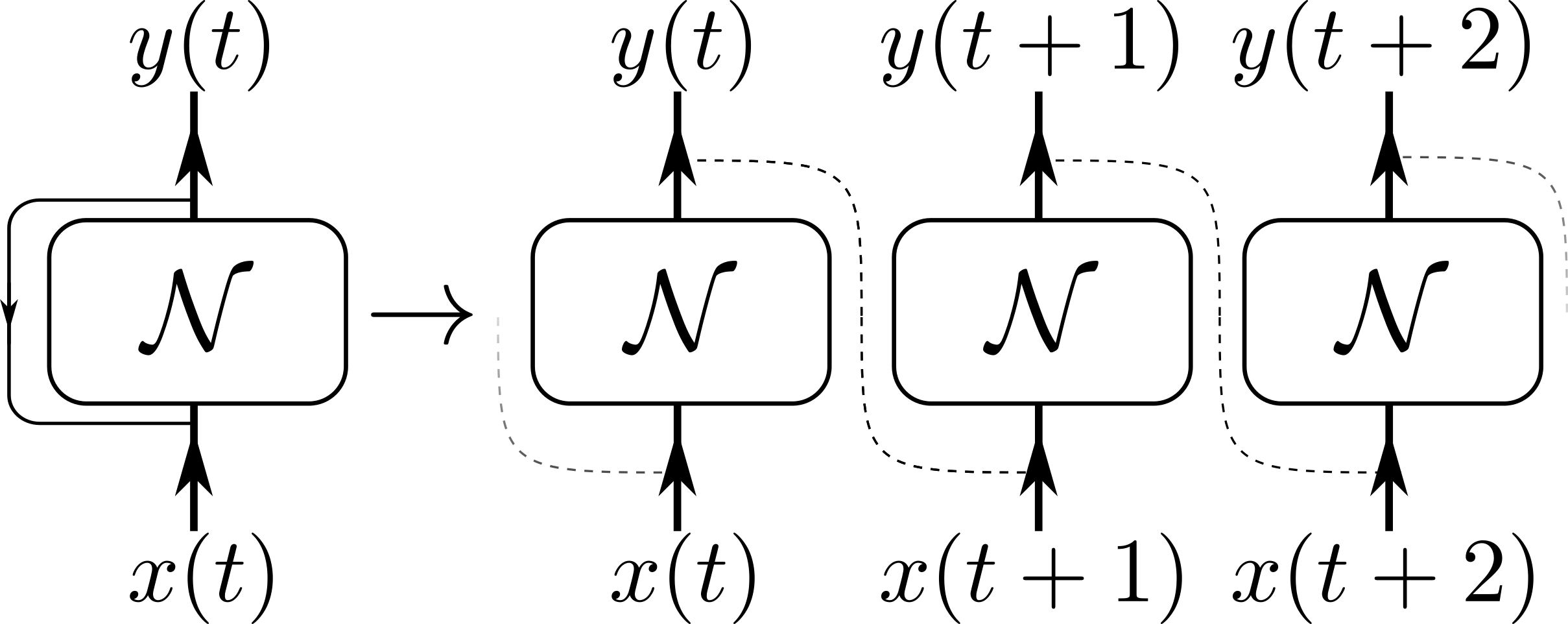}
        \caption{\textbf{Unrolling a recurrent network}. On the left a (subpart of a) neural network $\mathcal{N}$ is shown with output feeding back into input, making it into a recurennt neural network. On the right the unrolled version of the same network is shown, detailing that the output at step $t$ is fed back as an input for time step $t+1$. The recurrent connections have their own weights that are optimized during training.
        \label{fig:RNN}}
    \end{center}
\end{figure}

\emph{Recurrent networks} - Because we wish to be able to capture non-equal-time correlations in the magnetization traces, we choose to train a recurrent neural network (RNN) to distinguish dynamical regimes.
A recurrent neural network is a neural network in which one or multiple outputs are fed back into the network as inputs, as illustrated in Fig.~\ref{fig:RNN}.
Such a recurrence creates a feedback loop that allows information that was fed into the network to persist in a self-consistent manner.
This is ideal for analyzing sequences in which the value at a particular point of that sequence may depend on the previous entries. Consequently, RNNs are well suited for dealing with sequential data or other types of data for which a kind of `memory' or temporal dependence is beneficial. 

It is particularly useful to introduce the idea of `unrolling' a recurrent part of a network. In Figure~\ref{fig:RNN} we show a (subsection of) a neural network $\mathcal{N}$ with inputs $x(t)$ and outputs $y(t)$, the latter being fed back into the inputs. We think of $t$ here as a discrete parameter, such that inputs and outputs are computed at timesteps $t$, $t+1$, etcetera. The feedback should now be understood such that at timestep $t$, the network receives both $x(t)$ and $y(t-1)$ as its inputs, and produces $y(t)$ from them via an intermediate step. This is most easily visualized by the unrolled network shown in Fig.~\ref{fig:RNN}. Namely, the network keeps track of an internal state $h(t)$, which is updated according to $h(t) = f(h(t-1),x(t))$. The function $f$ represents the free parameters that we wish to learn by training the network. Given $h(t)$, the output $y(t) = g(h(t))$ is computed via another learnable function.  The training of such a network is done in a supervised manner identical to the standard feed-forward networks, except that it can be thought of as done `unrolled layer' by `unrolled layer'. There are various choices for the functions $f$ and $g$ introduced above, and we so-called Long Short Term Memory networks (LSTMs) ~\cite{HochreiterSchmidhuber}. We expect that the recurrence allows the network to build a better model governing the dynamics, which helps it in the task of classifying the inputs.

\emph{Blanking} - Since training the recurrent network requires labeled data (it is a supervised method), we use physics intuition to label the data only in the extremities of the phase space we consider, i.e. in the limits where we are confident about the physics of the system. The network is trained only in these regimes, and hence instead of the network seeing all the data we effectively `blank out' outside of the known limits. This blanking tests the network's ability to extract the underlying essential model of the data from these limits, and apply it to unseen data as a form of generalization. Care must be taken also that one supplies the network with enough and representative data such that a consistent model can, at least in princple, be extracted. As an important check we have tested that the predictions of the network are insensitive to adding slightly more or slightly less labeled data at the extremities (i.e. by shrinking or enlarging the blanked out region); that the network's confidence is correlated with its accuracy~\cite{Calibration}, and that the network assigns a confused output to phases it had not encountered during training (see Supplemental Material).

Additionally, one must check for and prevent the possibility of the network learning examples by heart (i.e. overfitting). We will employ dropout~\cite{Srivastava2014} and weight decay ($l_2$ regularization) to do so. We remark that empirically for models with disorder the many realizations and their variety even for a given disorder strength seem to already build in an inherent robustness against overfitting. The actual training of the network is done by minimizing the cross-entropy using the Adam optimizer~\cite{Kingma2015}. Additionally, we remark that the usual test-set validation can not be performed in the blanked region, since the network is not trained there.

Given the number $n$ of regions in which we know the physics (i.e. the number of expected phases), our networks are constructed with a softmax output layer with $n$ neurons. Thus, the networks take a sequence of magnetizations and output a probability distribution $\mathbf{p} = (p_1,\ldots,p_n)$ over the $n$ phases. This distribution describes the probability that the network assigns for the input sequence to belong to each of the phases $1,\ldots,n$. We then measure the confusion (uncertainty) of the network by examining the reduced distribution on the two most likely phases. Namely, assuming the probabilities are ordered by decreasing magnitudes $(p_1\geq p_2\geq \ldots)$, we define the confusion as $C= -\log_2( p_1 / (p_1+p_2) )$. The confusion $C$ vanishes when the network confidently predicts a specific phase ($p_1=1$, $p_2=0$), and it takes the maximal value of unity whenever the network cannot decide between two or more phases ($p_1=p_2$). Whenever the network changes its prediction from one phase to another at a certain value of an underlying parameter, the peak in $C$ surrounding this value can hence indicate the corresponding transition region.

\begin{figure*}[t!]
    \begin{center}
        \includegraphics[width=1.0\textwidth]{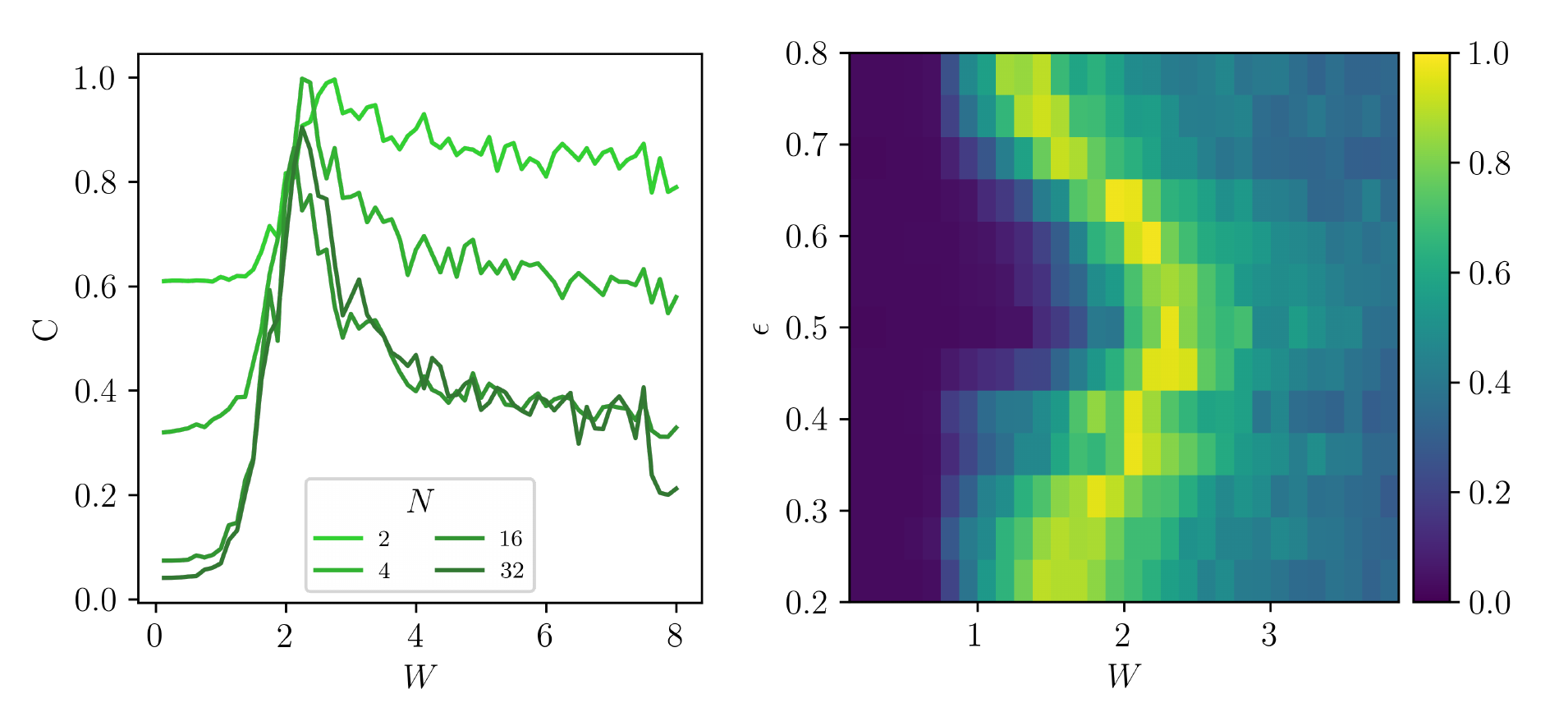}
        \caption{Detecting the MBL transition in the random-field Heisenberg model ~\ref{eq:H}. In the left panel, we show the dependence of the network's confusion $C$ on the number of LSTM neurons, $N$, for $\varepsilon=0.5$ and a fixed set of parameters: dropout $0.2$, $l_2 = 0.01$, batchsize of $64$ and $25$ training epochs. The right panel shows the resulting phase diagram (the colorbar represents the confusion $C$) in the $\varepsilon$ versus $W$ plane, obtained with $N=32$ and averaged over $10$ re-trainings.
        \label{fig:MBL}}
    \end{center}
\end{figure*}
\emph{MBL transition} - We consider the random-field Heisenberg model \cite{Pal2010}:
\begin{align}
 	H = \sum_{i} J\bm{S}_i\cdot\bm{S}_{i+1} + w_i S^z_i.
\label{eq:H}
\end{align}
The length of the chain is given by $L$, and the on-site disorders $w_i$ are drawn independently and uniformly from the interval $[-W,W]$. This Hamiltonian exhibits a transition between a delocalized and a many-body localized state at a critical disorder strength that depends on the energy density of the state under consideration~\cite{Pal2010,Luitz2015,Serbyn2015,Bera2015}. The dynamics of initial product states of spin polarization differs substantially between the two phases: while spins in the many-body localized phase retain long-term correlation with their initial configuration, in the delocalized phase this correlation is lost on much shorter timescales as expected from an ergodic system ~\cite{Iyer2013,Gopalakrishnan2016,Schreiber2015,Luitz2016}. In what follows we will be considering the dynamics of initial states that evolve in time under the Hamiltonian of Eq.~\ref{eq:H}, by performing exact time evolution on systems of size $L = 20$.

For the purpose of obtaining a phase diagram, we probe dynamics at various energy densities. Similarly to Ref.~\cite{Luitz2015}, we measure energy density by a parameter $\varepsilon$ interpolating between the minimal and maximal eigenenergies $E_0$, $E_{max}$ of each disorder realization. For each disorder realization we calculate $E_0$, $E_{max}$, and pick the product state in the $S_z$ basis ($\ket{\uparrow,\uparrow,\downarrow,\uparrow,\dots}$ etc.) whose energy expectation value is closest to $E=E_0+\varepsilon(E_{max}-E_0)$. We numerically evolve this initial state in time and measure $\langle \sigma^z_i \rangle (t)$ for each of the spins. 

The input to our networks therefore consists of these magnetization time-traces from $t = 0$ to $t = 500$, which we sample at $50$ equally spaced points, and hence is of shape $(L,50)$ for each disorder realization. At all energy densities considered we assume that the weak disorder regime ($W \leq 0.5$) is ergodic whilst the strong disorder regime ($W \geq 7.5$) is many-body localized. We therefore train the network on magnetization traces from these two extreme regimes. At low disorder these traces are labelled by a label $\mathbf{p} = (1,0)$, and at high disorder the label assigned is $\mathbf{p} = (0,1)$. Any data for disorder strengths in the interval $W \in [0.5,7.5]$ is therefore blanked out.

We fix the network architecture to have a single hidden layer of $N$ LSTM neurons with a dropout rate of $0.2$ and $l_2$ regularization of $0.01$, followed by a softmax layer to output a probability of the input being ergodic or non-ergodic. In the results below, we have re-trained the network $k=10$ times with identical parameters but different initial conditions. The results are averaged over these training cases.

We analyze the dependence of the output on the number $N$ of LSTM units in the left panel of Fig.~\ref{fig:MBL}, and find that with $32$ neurons we are able to converge the results for fixed batchsize $64$ and $25$ epochs. This training was done on the $\varepsilon = 0.5$ data, and uses the confidence enhancement introduced in~\cite{Schindler2017}. In order to gain a better understanding of what the LSTM neurons are doing, we analyze the case of a single LSTM neuron trained on a single-spin subsystem in the Supplemental Material. To obtain the phase diagram, we repeat the training process over the 2-dimensional parameter space of energy density ($13$ values equally spaced between $\varepsilon = 0.2$ and $\varepsilon = 0.8$) and disorder strength ($64$ values equally spaced between $W = 0.125$ and $W = 8$), with $50$ disorder realizations for each point. The obtained phase diagram is shown in the right panel of Fig.~\ref{fig:MBL}, and shows good agreement with the phase diagram obtained from static entanglement spectra in Ref.~\cite{Schindler2017}.

\emph{Time Crystals} - Next, we consider the following binary Floquet Hamiltonian acting on a one-dimensional spin-1/2 chain:
\begin{align} H =
\begin{cases}
  (g-\epsilon) \sum_{i} \sigma_i^x, & 0<t<T_1 \\
  \sum_{i}J_i\sigma_i^z\sigma_{i+1}^z + B_i^z\sigma_i^z, & T_1<t<T_2.
\end{cases}
\label{eq:H}
\end{align}

Where $J_i$, $B_i$ are random variables distributed independently and uniformly in the interval $[0, 0.5]$, $g$ is fixed to $\pi/2$, and $T_1 + T_2 = T$. This is a slight variation of the model studied in \cite{Yao2017}, where we took a different distribution for the bond terms $J_i$. To the best of our knowledge, our exact model has not been studied before, and its phase diagram has not been mapped out. Moreover, it serves as a case where a phase is inherently dynamical and cannot be studied in a static setting.

We are interested in the effect of the driving parameter $\epsilon$ on the resulting phase of the system. A guideline for the phases is provided through the long-time imbalance, $\mathcal{I}(t)$, defined as:
\begin{align}
	\mathcal{I}(t) = \frac{1}{L} \mathbf{m}(t)\cdot\mathbf{m}(0),
\end{align}
where the $i$-th component of $\mathbf{m}(t)$ is the expectation value of $\sigma_i^z$ at time $t$. This definition of the imbalance is the direct generalization of that typically used when the initial state is only taken to be one with a charge-density-wave ordering~\cite{Iyer2013,Gopalakrishnan2016,Schreiber2015,Luitz2016}.

\begin{figure}[t!]
    \begin{center}
        \includegraphics[width=0.5\textwidth]{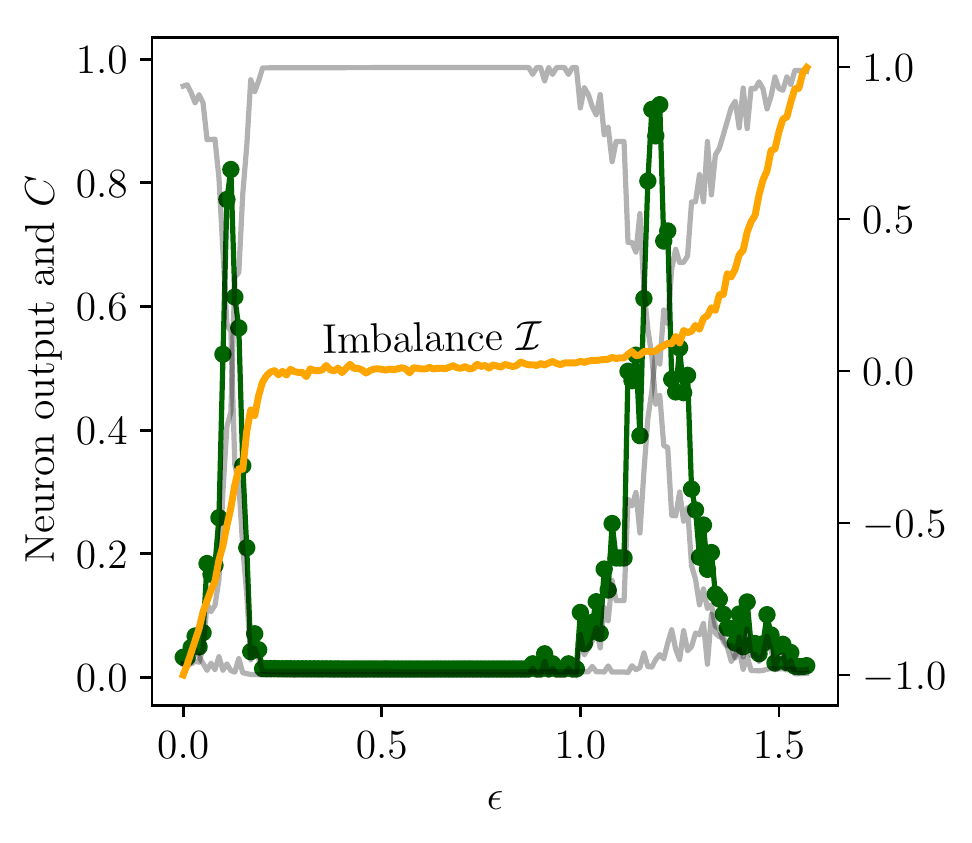}
        \caption{A recurrent neural network distinguishes between three dynamical phases of a time-dependent model, after being trained on example curves $\mathbf{m}(t)$ at $\epsilon = 0, 0.7$ and $\pi/2$. The gray curves show the outputs of the three neurons assigned to recognize each of the three phases (time-crystalline, Floquet-ergodic and Floquet-MBL). In green (with dots) the confusion $C$ of the network is shown, indicating two transition points between these phases. In orange we show the long-time imbalance $\mathcal{I}(t)$ measured at an odd driving period, taking a negative value in the time-crystalline phase; a vanishing value in the Floquet-ergodic phase; and a positive value in the Floquet-MBL phase. The phase boundaries extracted by the network are consistent with, and seem more sharp than, the imbalance.
        \label{fig:TC}}
    \end{center}
\end{figure}

The long-time imbalance shows three distinct behaviours as a function of the driving parameter $\epsilon$ (orange line in Fig. ~\ref{fig:TC}). If $\epsilon = \pi/2$, the drive term is just an identity operator and the system is governed by a many-body localized Hamiltonian. Subsequently, for $\epsilon$ sufficiently close to  $\pi/2$, the imbalance retains a value close to its initial one, indicating a trivial Floquet-MBL phase. For intermediate values of $\epsilon$, the long-time imbalance vanishes, indicating a transition to a Floquet-ergodic phase. Interestingly, below a critical value of $\epsilon$, the long-term imbalance retains a value that is close to its initial one in magnitude, but flips sign every driving period. In this regime the system's response is periodic in $2T$ rather than $T$, leading to the nomenclature `time crystal'.

We proceed with training a RNN on time traces of $\mathbf{m}(t)$ identically to the case of the previously discussed MBL system, apart from having three regions in phase space where we train the network instead of two. Namely, for $\epsilon$ close to $0$ we assign the time-crystalline label, for $\epsilon \approx 0.7$ we assign the Floquet-ergodic label and for $\epsilon = \pi/2$ we assign the Floquet-MBL label. We again use $32$ LSTM units, dropout $0.2$ and $l_2 = 0.01$ with Adam optimization. When evaluated on a data-set with many more $\epsilon$ available, the resulting 1D phase diagram is shown in Fig.~\ref{fig:TC} (green and gray lines). \

\emph{Discussion and Prospects} - The main point considered in this work was the study of \emph{dynamics} using machine learning methods, and doing so using \emph{experimentally} available measurements. We employed recurrent neural networks, rather than their non-recurrent variants. There are multiple motivations, apart from the input being sequences, for using such an approach over more common non-recurrent feed-forward networks. First, since the data is fed into the network one time-step at a time, the number of network parameters does not scale with the number of time-steps. This also means that the same recurrent network can be easily trained on various lengths of data. In contrast, a regular feed-forward network would need to be input with all of the data at once, leading to a large initial input layer compatible with a fixed input length. We have studied whether the use of recurrent neurons provides a more direct way of extracting what feature of the data the neurons use to output their guess. By training one or multiple LSTM units on single magnetization curves, it is possible to identify neuron behavior (see Supplemental Material). We speculate that it might be possible to extract from these results a `dynamical order parameter', which takes the full magnetization traces into account.

Using the networks, we constructed dynamical phase diagrams for the MBL transition and a driven model featuring a time-crystalline phase, thereby circumventing the need to manually construct a threshold criterion or dynamical order parameter for locating the phase boundary. Rather, such a threshold was automatically determined from the data. By considering the LSTM neuron outputs for the single spin case (see Supplemental Material), we were able to gain some understanding of the behaviour that the network latches onto.

We emphasize that obtaining a phase boundary from data can hence only be as accurate as the available data. The boundary we obtain for the MBL transition is at a slightly lower disorder strength than that of the exact diagonalization results in Ref.~\cite{Luitz2015}, but agrees well with that obtained using the machine learned entanglement spectra of Ref.~\cite{Schindler2017}. The alternative of finding a non-machine learned proxy to serve as an indicator, such as the imbalance for the MBL transition (see Supplemental Material), can be ambiguous. If sufficient data are available, we expect consistency of that data to be the judge of where the transition happens. It may be possible to use the same criterion in a feedback system between a machine learning algorithm and experiments, where measurements performed on the experiment are chosen to improve the phase boundary.

For the MBL transition in particular, we mention that a more detailed investigation should also take into account the possibility of a Griffiths phase, possibly showing up as a region where the network prediction is increasingly uncertain as system size increases. Such a finite-size scaling can indeed be succesfully attempted using machine learned data~\cite{Carrasquilla2016,Beach2017}, and provides a useful and interesting alternative for locating a phase boundary.

Being able to train recurrent neural networks on time-traces of data poses the question of whether such methods can be used to enhance the prediction of dynamics, i.e. in numerical time evolution simulations. Such questions are being actively adressed in order to provide accurate control over e.g. single qubits in decohering environments and noisy measurements~\cite{Gupta2017}.


\vspace{-0.3cm}
\begin{thebibliography}{71}
\expandafter\ifx\csname natexlab\endcsname\relax\def\natexlab#1{#1}\fi
\expandafter\ifx\csname bibnamefont\endcsname\relax
  \def\bibnamefont#1{#1}\fi
\expandafter\ifx\csname bibfnamefont\endcsname\relax
  \def\bibfnamefont#1{#1}\fi
\expandafter\ifx\csname citenamefont\endcsname\relax
  \def\citenamefont#1{#1}\fi
\expandafter\ifx\csname url\endcsname\relax
  \def\url#1{\texttt{#1}}\fi
\expandafter\ifx\csname urlprefix\endcsname\relax\def\urlprefix{URL }\fi
\providecommand{\bibinfo}[2]{#2}
\providecommand{\eprint}[2][]{\url{#2}}

\bibitem[{\citenamefont{Carrasquilla and Melko}(2017)}]{Carrasquilla2016}
\bibinfo{author}{\bibfnamefont{J.}~\bibnamefont{Carrasquilla}}
  \bibnamefont{and} \bibinfo{author}{\bibfnamefont{R.~G.} \bibnamefont{Melko}},
  \bibinfo{journal}{Nat. Phys.} \textbf{\bibinfo{volume}{13}},
  \bibinfo{pages}{431} (\bibinfo{year}{2017}), ISSN \bibinfo{issn}{1745-2473},
  \eprint{1605.01735}, \urlprefix\url{http://dx.doi.org/10.1038/nphys4035}.

\bibitem[{\citenamefont{Wang}(2016)}]{Wang2016}
\bibinfo{author}{\bibfnamefont{L.}~\bibnamefont{Wang}}, \bibinfo{journal}{Phys.
  Rev. B} \textbf{\bibinfo{volume}{94}}, \bibinfo{pages}{195105}
  (\bibinfo{year}{2016}), ISSN \bibinfo{issn}{2469-9950}, \eprint{1606.00318},
  \urlprefix\url{https://link.aps.org/doi/10.1103/PhysRevB.94.195105}.

\bibitem[{\citenamefont{van Nieuwenburg et~al.}(2017)\citenamefont{van
  Nieuwenburg, Liu, and Huber}}]{VanNieuwenburg2017}
\bibinfo{author}{\bibfnamefont{E.~P.~L.} \bibnamefont{van Nieuwenburg}},
  \bibinfo{author}{\bibfnamefont{Y.-H.} \bibnamefont{Liu}}, \bibnamefont{and}
  \bibinfo{author}{\bibfnamefont{S.~D.} \bibnamefont{Huber}},
  \bibinfo{journal}{Nat. Phys.} \textbf{\bibinfo{volume}{13}},
  \bibinfo{pages}{435} (\bibinfo{year}{2017}), ISSN \bibinfo{issn}{1745-2473},
  \eprint{1610.02048}, \urlprefix\url{http://arxiv.org/abs/1610.02048
  http://www.nature.com/doifinder/10.1038/nphys4037}.

\bibitem[{\citenamefont{Broecker et~al.}(2017)\citenamefont{Broecker, Assaad,
  and Trebst}}]{Broecker2017}
\bibinfo{author}{\bibfnamefont{P.}~\bibnamefont{Broecker}},
  \bibinfo{author}{\bibfnamefont{F.}~\bibnamefont{Assaad}}, \bibnamefont{and}
  \bibinfo{author}{\bibfnamefont{S.}~\bibnamefont{Trebst}},
  \bibinfo{journal}{arXiv:1707.00663}  (\bibinfo{year}{2017}).

\bibitem[{\citenamefont{Wetzel and Scherzer}(2017)}]{Wetzel2017}
\bibinfo{author}{\bibfnamefont{S.~J.} \bibnamefont{Wetzel}} \bibnamefont{and}
  \bibinfo{author}{\bibfnamefont{M.}~\bibnamefont{Scherzer}},
  \bibinfo{journal}{arXiv:1705.05582}  (\bibinfo{year}{2017}),
  \eprint{1705.05582}, \urlprefix\url{http://arxiv.org/abs/1705.05582
  https://arxiv.org/pdf/1705.05582.pdf}.

\bibitem[{\citenamefont{Wetzel}(2017)}]{Wetzel2017a}
\bibinfo{author}{\bibfnamefont{S.~J.} \bibnamefont{Wetzel}},
  \bibinfo{journal}{arXiv:1703.02435}  (\bibinfo{year}{2017}),
  \eprint{1703.02435}, \urlprefix\url{http://arxiv.org/abs/1703.02435}.

\bibitem[{\citenamefont{Liu and van Nieuwenburg}(2017)}]{LiuNieuwenburg2017}
\bibinfo{author}{\bibfnamefont{Y.-H.} \bibnamefont{Liu}} \bibnamefont{and}
  \bibinfo{author}{\bibfnamefont{E.~P.~L.} \bibnamefont{van Nieuwenburg}}
  (\bibinfo{year}{2017}), \eprint{1706.08111},
  \urlprefix\url{http://arxiv.org/abs/1706.08111}.

\bibitem[{\citenamefont{Ch'ng et~al.}(2016)\citenamefont{Ch'ng, Carrasquilla,
  Melko, and Khatami}}]{Chng2016}
\bibinfo{author}{\bibfnamefont{K.}~\bibnamefont{Ch'ng}},
  \bibinfo{author}{\bibfnamefont{J.}~\bibnamefont{Carrasquilla}},
  \bibinfo{author}{\bibfnamefont{R.~G.} \bibnamefont{Melko}}, \bibnamefont{and}
  \bibinfo{author}{\bibfnamefont{E.}~\bibnamefont{Khatami}},
  \bibinfo{journal}{arXiv:1609.02552}  (\bibinfo{year}{2016}),
  \eprint{1609.02552}, \urlprefix\url{http://arxiv.org/abs/1609.02552}.

\bibitem[{\citenamefont{Ch'ng et~al.}(2018)\citenamefont{Ch'ng, Vazquez, and
  Khatami}}]{Chng2018}
\bibinfo{author}{\bibfnamefont{K.}~\bibnamefont{Ch'ng}},
  \bibinfo{author}{\bibfnamefont{N.}~\bibnamefont{Vazquez}}, \bibnamefont{and}
  \bibinfo{author}{\bibfnamefont{E.}~\bibnamefont{Khatami}},
  \bibinfo{journal}{Phys. Rev. E} \textbf{\bibinfo{volume}{97}},
  \bibinfo{pages}{013306} (\bibinfo{year}{2018}),
  \urlprefix\url{https://link.aps.org/doi/10.1103/PhysRevE.97.013306}.

\bibitem[{\citenamefont{Broecker et~al.}(2016)\citenamefont{Broecker,
  Carrasquilla, Melko, and Trebst}}]{Broecker2016}
\bibinfo{author}{\bibfnamefont{P.}~\bibnamefont{Broecker}},
  \bibinfo{author}{\bibfnamefont{J.}~\bibnamefont{Carrasquilla}},
  \bibinfo{author}{\bibfnamefont{R.~G.} \bibnamefont{Melko}}, \bibnamefont{and}
  \bibinfo{author}{\bibfnamefont{S.}~\bibnamefont{Trebst}},
  \bibinfo{journal}{arXiv:1608.07848}  (\bibinfo{year}{2016}),
  \eprint{1608.07848}, \urlprefix\url{http://arxiv.org/abs/1608.07848}.

\bibitem[{\citenamefont{Schindler et~al.}(2017)\citenamefont{Schindler,
  Regnault, and Neupert}}]{Schindler2017}
\bibinfo{author}{\bibfnamefont{F.}~\bibnamefont{Schindler}},
  \bibinfo{author}{\bibfnamefont{N.}~\bibnamefont{Regnault}}, \bibnamefont{and}
  \bibinfo{author}{\bibfnamefont{T.}~\bibnamefont{Neupert}}, \href{http://link.aps.org/doi/10.1103/PhysRevB.95.245134}
  {\bibfield {journal} \bibinfo{journal}{Physical Review B} \textbf{\bibinfo{volume}{95}},
  \bibinfo{pages}{245134} (\bibinfo{year}{2017}) }	

\bibitem[{\citenamefont{Ohtsuki and Ohtsuki}(2016)}]{Ohtsuki2016}
\bibinfo{author}{\bibfnamefont{T.}~\bibnamefont{Ohtsuki}} \bibnamefont{and}
  \bibinfo{author}{\bibfnamefont{T.}~\bibnamefont{Ohtsuki}},
  \bibinfo{journal}{J. Phys. Soc. Japan} \textbf{\bibinfo{volume}{85}},
  \bibinfo{pages}{123706} (\bibinfo{year}{2016}), ISSN
  \bibinfo{issn}{0031-9015}, \eprint{1610.00462},
  \urlprefix\url{http://journals.jps.jp/doi/10.7566/JPSJ.85.123706}.

\bibitem[{\citenamefont{Arsenault et~al.}(2014)\citenamefont{Arsenault,
  Lopez-Bezanilla, von Lilienfeld, and Millis}}]{Arsenault2014a}
\bibinfo{author}{\bibfnamefont{L.-F.} \bibnamefont{Arsenault}},
  \bibinfo{author}{\bibfnamefont{A.}~\bibnamefont{Lopez-Bezanilla}},
  \bibinfo{author}{\bibfnamefont{O.~A.} \bibnamefont{von Lilienfeld}},
  \bibnamefont{and} \bibinfo{author}{\bibfnamefont{A.~J.}
  \bibnamefont{Millis}}, \bibinfo{journal}{Phys. Rev. B}
  \textbf{\bibinfo{volume}{90}}, \bibinfo{pages}{155136}
  (\bibinfo{year}{2014}), ISSN \bibinfo{issn}{1098-0121}, \eprint{1408.1143},
  \urlprefix\url{https://link.aps.org/doi/10.1103/PhysRevB.90.155136}.

\bibitem[{\citenamefont{Arsenault et~al.}(2015)\citenamefont{Arsenault, von
  Lilienfeld, and Millis}}]{Arsenault2015}
\bibinfo{author}{\bibfnamefont{L.-F.} \bibnamefont{Arsenault}},
  \bibinfo{author}{\bibfnamefont{O.~A.} \bibnamefont{von Lilienfeld}},
  \bibnamefont{and} \bibinfo{author}{\bibfnamefont{A.~J.}
  \bibnamefont{Millis}}, \bibinfo{journal}{arXiv:1506.08858}
  (\bibinfo{year}{2015}), \eprint{1506.08858},
  \urlprefix\url{http://arxiv.org/abs/1506.08858v1}.

\bibitem[{\citenamefont{Beach et~al.}(2017)\citenamefont{Beach, Golubeva, and
  Melko}}]{Beach2017}
\bibinfo{author}{\bibfnamefont{M.~J.} \bibnamefont{Beach}},
  \bibinfo{author}{\bibfnamefont{A.}~\bibnamefont{Golubeva}}, \bibnamefont{and}
  \bibinfo{author}{\bibfnamefont{R.~G.} \bibnamefont{Melko}},
  \bibinfo{journal}{arXiv:1710.09842}  (\bibinfo{year}{2017}),
  \eprint{1710.09842}, \urlprefix\url{https://arxiv.org/abs/1710.09842}.

\bibitem[{\citenamefont{Yoshioka et~al.}(2017)\citenamefont{Yoshioka, Akagi,
  and Katsura}}]{Yoshioka2017}
\bibinfo{author}{\bibfnamefont{N.}~\bibnamefont{Yoshioka}},
  \bibinfo{author}{\bibfnamefont{Y.}~\bibnamefont{Akagi}}, \bibnamefont{and}
  \bibinfo{author}{\bibfnamefont{H.}~\bibnamefont{Katsura}},
  \bibinfo{journal}{arXiv e-prints}  (\bibinfo{year}{2017}),
  \eprint{1709.05790}, \urlprefix\url{http://arxiv.org/abs/1709.05790}.

\bibitem[{\citenamefont{Venderley et~al.}(2017)\citenamefont{Venderley,
  Khemani, and Kim}}]{Venderley2017}
\bibinfo{author}{\bibfnamefont{J.}~\bibnamefont{Venderley}},
  \bibinfo{author}{\bibfnamefont{V.}~\bibnamefont{Khemani}}, \bibnamefont{and}
  \bibinfo{author}{\bibfnamefont{E.-A.} \bibnamefont{Kim}}
  (\bibinfo{year}{2017}), \eprint{arXiv:1711.00020}.

\bibitem[{\citenamefont{Carleo and Troyer}(2017)}]{Carleo2016}
\bibinfo{author}{\bibfnamefont{G.}~\bibnamefont{Carleo}} \bibnamefont{and}
  \bibinfo{author}{\bibfnamefont{M.}~\bibnamefont{Troyer}},
  \bibinfo{journal}{Science} \textbf{\bibinfo{volume}{355}},
  \bibinfo{pages}{602} (\bibinfo{year}{2017}), ISSN \bibinfo{issn}{0036-8075},
  \eprint{1606.02318},
  \urlprefix\url{http://www.sciencemag.org/lookup/doi/10.1126/science.aag2302}.

\bibitem[{\citenamefont{Schmitt and Heyl}(2017)}]{Schmitt2017}
\bibinfo{author}{\bibfnamefont{M.}~\bibnamefont{Schmitt}} \bibnamefont{and}
  \bibinfo{author}{\bibfnamefont{M.}~\bibnamefont{Heyl}}
  (\bibinfo{year}{2017}), \eprint{1707.06656},
  \urlprefix\url{http://arxiv.org/abs/1707.06656}.

\bibitem[{\citenamefont{Cai}(2017)}]{Cai2017}
\bibinfo{author}{\bibfnamefont{Z.}~\bibnamefont{Cai}},
  \bibinfo{journal}{arXiv:1704.05148}  (\bibinfo{year}{2017}),
  \eprint{1704.05148}, \urlprefix\url{http://arxiv.org/abs/1704.05148}.

\bibitem[{\citenamefont{Huang and Moore}(2017)}]{Huang2017}
\bibinfo{author}{\bibfnamefont{Y.}~\bibnamefont{Huang}} \bibnamefont{and}
  \bibinfo{author}{\bibfnamefont{J.~E.} \bibnamefont{Moore}}
  (\bibinfo{year}{2017}), \eprint{1701.06246},
  \urlprefix\url{http://arxiv.org/abs/1701.06246}.

\bibitem[{\citenamefont{Deng et~al.}(2016)\citenamefont{Deng, Li, and
  Sarma}}]{Deng2016}
\bibinfo{author}{\bibfnamefont{D.-L.} \bibnamefont{Deng}},
  \bibinfo{author}{\bibfnamefont{X.}~\bibnamefont{Li}}, \bibnamefont{and}
  \bibinfo{author}{\bibfnamefont{S.~D.} \bibnamefont{Sarma}},
  \bibinfo{journal}{arXiv:1609.09060}  (\bibinfo{year}{2016}),
  \eprint{1609.09060}, \urlprefix\url{http://arxiv.org/abs/1609.09060}.

\bibitem[{\citenamefont{Nomura et~al.}(2017)\citenamefont{Nomura, Darmawan,
  Yamaji, and Imada}}]{Nomura2017}
\bibinfo{author}{\bibfnamefont{Y.}~\bibnamefont{Nomura}},
  \bibinfo{author}{\bibfnamefont{A.}~\bibnamefont{Darmawan}},
  \bibinfo{author}{\bibfnamefont{Y.}~\bibnamefont{Yamaji}}, \bibnamefont{and}
  \bibinfo{author}{\bibfnamefont{M.}~\bibnamefont{Imada}}
  (\bibinfo{year}{2017}), \eprint{arXiv:1709.06475v1},
  \urlprefix\url{https://arxiv.org/pdf/1709.06475.pdf}.

\bibitem[{\citenamefont{Deng et~al.}(2017)\citenamefont{Deng, Li, and
  Sarma}}]{Deng2017}
\bibinfo{author}{\bibfnamefont{D.-L.} \bibnamefont{Deng}},
  \bibinfo{author}{\bibfnamefont{X.}~\bibnamefont{Li}}, \bibnamefont{and}
  \bibinfo{author}{\bibfnamefont{S.~D.} \bibnamefont{Sarma}}
  (\bibinfo{year}{2017}), \eprint{1701.04844},
  \urlprefix\url{http://arxiv.org/abs/1701.04844}.

\bibitem[{\citenamefont{Gao and Duan}(2017)}]{Gao2017}
\bibinfo{author}{\bibfnamefont{X.}~\bibnamefont{Gao}} \bibnamefont{and}
  \bibinfo{author}{\bibfnamefont{L.-M.} \bibnamefont{Duan}},
  \bibinfo{journal}{arXiv:1701.05039}  (\bibinfo{year}{2017}),
  \eprint{1701.05039}, \urlprefix\url{http://arxiv.org/abs/1701.05039}.

\bibitem[{\citenamefont{Torlai et~al.}(2017)\citenamefont{Torlai, Mazzola,
  Carrasquilla, Troyer, Melko, and Carleo}}]{Torlai2017}
\bibinfo{author}{\bibfnamefont{G.}~\bibnamefont{Torlai}},
  \bibinfo{author}{\bibfnamefont{G.}~\bibnamefont{Mazzola}},
  \bibinfo{author}{\bibfnamefont{J.}~\bibnamefont{Carrasquilla}},
  \bibinfo{author}{\bibfnamefont{M.}~\bibnamefont{Troyer}},
  \bibinfo{author}{\bibfnamefont{R.}~\bibnamefont{Melko}}, \bibnamefont{and}
  \bibinfo{author}{\bibfnamefont{G.}~\bibnamefont{Carleo}},
  \bibinfo{journal}{arXiv:1703.05334}  (\bibinfo{year}{2017}),
  \eprint{1703.05334}, \urlprefix\url{http://arxiv.org/abs/1703.05334}.

\bibitem[{\citenamefont{Mehta and Schwab}(2014)}]{Mehta2014}
\bibinfo{author}{\bibfnamefont{P.}~\bibnamefont{Mehta}} \bibnamefont{and}
  \bibinfo{author}{\bibfnamefont{D.~J.} \bibnamefont{Schwab}},
  \bibinfo{journal}{arXiv:1410.3831}  (\bibinfo{year}{2014}), ISSN
  \bibinfo{issn}{01406736}, \eprint{1410.3831},
  \urlprefix\url{http://arxiv.org/abs/1410.3831}.

\bibitem[{\citenamefont{Koch-Janusz and Ringel}(2018)}]{Koch-Janusz2017}
\bibinfo{author}{\bibfnamefont{M.}~\bibnamefont{Koch-Janusz}} \bibnamefont{and}
  \bibinfo{author}{\bibfnamefont{Z.}~\bibnamefont{Ringel}},
  \bibinfo{journal}{Nature Physics}  (\bibinfo{year}{2018}),
  \eprint{1704.06279}, \urlprefix\url{https://doi.org/10.1038/s41567-018-0081-4
  https://doi.org/10.1038/s41567-018-0081-4}.

\bibitem[{\citenamefont{Li and Wang}(2018)}]{Wang2018}
\bibinfo{author}{\bibfnamefont{S.-H.} \bibnamefont{Li}} \bibnamefont{and}
  \bibinfo{author}{\bibfnamefont{L.}~\bibnamefont{Wang}}
  (\bibinfo{year}{2018}), \eprint{1802.02840},
  \urlprefix\url{http://arxiv.org/abs/1802.02840}.

\bibitem[{\citenamefont{Moessner and Sondhi}(2017)}]{Moessner2017}
\bibinfo{author}{\bibfnamefont{R.}~\bibnamefont{Moessner}} \bibnamefont{and}
  \bibinfo{author}{\bibfnamefont{S.~L.} \bibnamefont{Sondhi}},
  \bibinfo{journal}{Nature Physics} \textbf{\bibinfo{volume}{13}},
  \bibinfo{pages}{424 EP } (\bibinfo{year}{2017}),
  \urlprefix\url{http://dx.doi.org/10.1038/nphys4106}.

\bibitem[{\citenamefont{Lindner et~al.}(2011)\citenamefont{Lindner, Refael, and
  Galitski}}]{Lindner2011}
\bibinfo{author}{\bibfnamefont{N.~H.} \bibnamefont{Lindner}},
  \bibinfo{author}{\bibfnamefont{G.}~\bibnamefont{Refael}}, \bibnamefont{and}
  \bibinfo{author}{\bibfnamefont{V.}~\bibnamefont{Galitski}},
  \bibinfo{journal}{Nature Physics} \textbf{\bibinfo{volume}{7}},
  \bibinfo{pages}{490 EP } (\bibinfo{year}{2011}), \bibinfo{note}{article},
  \urlprefix\url{http://dx.doi.org/10.1038/nphys1926}.

\bibitem[{\citenamefont{Titum et~al.}(2016)\citenamefont{Titum, Berg, Rudner,
  Refael, and Lindner}}]{Titum2016}
\bibinfo{author}{\bibfnamefont{P.}~\bibnamefont{Titum}},
  \bibinfo{author}{\bibfnamefont{E.}~\bibnamefont{Berg}},
  \bibinfo{author}{\bibfnamefont{M.~S.} \bibnamefont{Rudner}},
  \bibinfo{author}{\bibfnamefont{G.}~\bibnamefont{Refael}}, \bibnamefont{and}
  \bibinfo{author}{\bibfnamefont{N.~H.} \bibnamefont{Lindner}},
  \bibinfo{journal}{Phys. Rev. X} \textbf{\bibinfo{volume}{6}},
  \bibinfo{pages}{021013} (\bibinfo{year}{2016}),
  \urlprefix\url{https://link.aps.org/doi/10.1103/PhysRevX.6.021013}.

\bibitem[{\citenamefont{Nathan et~al.}(2017)\citenamefont{Nathan, Rudner,
  Lindner, Berg, and Refael}}]{Nathan2017}
\bibinfo{author}{\bibfnamefont{F.}~\bibnamefont{Nathan}},
  \bibinfo{author}{\bibfnamefont{M.~S.} \bibnamefont{Rudner}},
  \bibinfo{author}{\bibfnamefont{N.~H.} \bibnamefont{Lindner}},
  \bibinfo{author}{\bibfnamefont{E.}~\bibnamefont{Berg}}, \bibnamefont{and}
  \bibinfo{author}{\bibfnamefont{G.}~\bibnamefont{Refael}},
  \bibinfo{journal}{Phys. Rev. Lett.} \textbf{\bibinfo{volume}{119}},
  \bibinfo{pages}{186801} (\bibinfo{year}{2017}),
  \urlprefix\url{https://link.aps.org/doi/10.1103/PhysRevLett.119.186801}.

\bibitem[{\citenamefont{Else et~al.}(2017)\citenamefont{Else, Bauer, and
  Nayak}}]{Else2017}
\bibinfo{author}{\bibfnamefont{D.~V.} \bibnamefont{Else}},
  \bibinfo{author}{\bibfnamefont{B.}~\bibnamefont{Bauer}}, \bibnamefont{and}
  \bibinfo{author}{\bibfnamefont{C.}~\bibnamefont{Nayak}},
  \bibinfo{journal}{Phys. Rev. X} \textbf{\bibinfo{volume}{7}},
  \bibinfo{pages}{011026} (\bibinfo{year}{2017}),
  \urlprefix\url{https://link.aps.org/doi/10.1103/PhysRevX.7.011026}.

\bibitem[{\citenamefont{Heyl}(2017)}]{Heyl2017}
\bibinfo{author}{\bibfnamefont{M.}~\bibnamefont{Heyl}},
  \emph{\bibinfo{title}{Dynamical quantum phase transitions: a review}}
  (\bibinfo{year}{2017}), \eprint{arXiv:1709.07461}.

\bibitem[{\citenamefont{von Keyserlingk and
  Sondhi}(2016)}]{vonKeyserlingk2016b}
\bibinfo{author}{\bibfnamefont{C.~W.} \bibnamefont{von Keyserlingk}}
  \bibnamefont{and} \bibinfo{author}{\bibfnamefont{S.~L.}
  \bibnamefont{Sondhi}}, \bibinfo{journal}{Phys. Rev. B}
  \textbf{\bibinfo{volume}{93}}, \bibinfo{pages}{245145}
  (\bibinfo{year}{2016}),
  \urlprefix\url{https://link.aps.org/doi/10.1103/PhysRevB.93.245145}.

\bibitem[{\citenamefont{Else and Nayak}(2016)}]{Else2016b}
\bibinfo{author}{\bibfnamefont{D.~V.} \bibnamefont{Else}} \bibnamefont{and}
  \bibinfo{author}{\bibfnamefont{C.}~\bibnamefont{Nayak}},
  \bibinfo{journal}{Phys. Rev. B} \textbf{\bibinfo{volume}{93}},
  \bibinfo{pages}{201103} (\bibinfo{year}{2016}),
  \urlprefix\url{https://link.aps.org/doi/10.1103/PhysRevB.93.201103}.

\bibitem[{\citenamefont{Po et~al.}(2016)\citenamefont{Po, Fidkowski, Morimoto,
  Potter, and Vishwanath}}]{Po2016}
\bibinfo{author}{\bibfnamefont{H.~C.} \bibnamefont{Po}},
  \bibinfo{author}{\bibfnamefont{L.}~\bibnamefont{Fidkowski}},
  \bibinfo{author}{\bibfnamefont{T.}~\bibnamefont{Morimoto}},
  \bibinfo{author}{\bibfnamefont{A.~C.} \bibnamefont{Potter}},
  \bibnamefont{and}
  \bibinfo{author}{\bibfnamefont{A.}~\bibnamefont{Vishwanath}},
  \bibinfo{journal}{Phys. Rev. X} \textbf{\bibinfo{volume}{6}},
  \bibinfo{pages}{041070} (\bibinfo{year}{2016}),
  \urlprefix\url{https://link.aps.org/doi/10.1103/PhysRevX.6.041070}.

\bibitem[{\citenamefont{Potter et~al.}(2016)\citenamefont{Potter, Morimoto, and
  Vishwanath}}]{Potter2016}
\bibinfo{author}{\bibfnamefont{A.~C.} \bibnamefont{Potter}},
  \bibinfo{author}{\bibfnamefont{T.}~\bibnamefont{Morimoto}}, \bibnamefont{and}
  \bibinfo{author}{\bibfnamefont{A.}~\bibnamefont{Vishwanath}},
  \bibinfo{journal}{Phys. Rev. X} \textbf{\bibinfo{volume}{6}},
  \bibinfo{pages}{041001} (\bibinfo{year}{2016}),
  \urlprefix\url{https://link.aps.org/doi/10.1103/PhysRevX.6.041001}.

\bibitem[{\citenamefont{Basko et~al.}(2006)\citenamefont{Basko, Aleiner, and
  Altshuler}}]{Basko2006}
\bibinfo{author}{\bibfnamefont{D.}~\bibnamefont{Basko}},
  \bibinfo{author}{\bibfnamefont{I.}~\bibnamefont{Aleiner}}, \bibnamefont{and}
  \bibinfo{author}{\bibfnamefont{B.}~\bibnamefont{Altshuler}},
  \bibinfo{journal}{Annals of Physics} \textbf{\bibinfo{volume}{321}},
  \bibinfo{pages}{1126 } (\bibinfo{year}{2006}), ISSN
  \bibinfo{issn}{0003-4916},
  \urlprefix\url{http://www.sciencedirect.com/science/article/pii/S0003491605002630}.

\bibitem[{\citenamefont{Oganesyan and Huse}(2007)}]{Oganesyan2007}
\bibinfo{author}{\bibfnamefont{V.}~\bibnamefont{Oganesyan}} \bibnamefont{and}
  \bibinfo{author}{\bibfnamefont{D.~A.} \bibnamefont{Huse}},
  \bibinfo{journal}{Phys. Rev. B} \textbf{\bibinfo{volume}{75}},
  \bibinfo{pages}{155111} (\bibinfo{year}{2007}),
  \urlprefix\url{https://link.aps.org/doi/10.1103/PhysRevB.75.155111}.

\bibitem[{\citenamefont{Nandkishore and Huse}(2015)}]{Nandkishore2015}
\bibinfo{author}{\bibfnamefont{R.}~\bibnamefont{Nandkishore}} \bibnamefont{and}
  \bibinfo{author}{\bibfnamefont{D.~A.} \bibnamefont{Huse}},
  \bibinfo{journal}{Annual Review of Condensed Matter Physics}
  \textbf{\bibinfo{volume}{6}}, \bibinfo{pages}{15} (\bibinfo{year}{2015}),
  \urlprefix\url{https://doi.org/10.1146/annurev-conmatphys-031214-014726}.

\bibitem[{\citenamefont{Abanin and Papić}(2017)}]{Abanin2017}
\bibinfo{author}{\bibfnamefont{D.~A.} \bibnamefont{Abanin}} \bibnamefont{and}
  \bibinfo{author}{\bibfnamefont{Z.}~\bibnamefont{Papić}},
  \bibinfo{journal}{Annalen der Physik} \textbf{\bibinfo{volume}{529}},
  \bibinfo{pages}{1700169} (\bibinfo{year}{2017}), ISSN
  \bibinfo{issn}{1521-3889}, \bibinfo{note}{1700169},
  \urlprefix\url{http://dx.doi.org/10.1002/andp.201700169}.

\bibitem[{\citenamefont{Else et~al.}(2016)\citenamefont{Else, Bauer, and
  Nayak}}]{Else2016}
\bibinfo{author}{\bibfnamefont{D.~V.} \bibnamefont{Else}},
  \bibinfo{author}{\bibfnamefont{B.}~\bibnamefont{Bauer}}, \bibnamefont{and}
  \bibinfo{author}{\bibfnamefont{C.}~\bibnamefont{Nayak}},
  \bibinfo{journal}{Phys. Rev. Lett.} \textbf{\bibinfo{volume}{117}},
  \bibinfo{pages}{090402} (\bibinfo{year}{2016}),
  \urlprefix\url{https://link.aps.org/doi/10.1103/PhysRevLett.117.090402}.

\bibitem[{\citenamefont{Khemani et~al.}(2016)\citenamefont{Khemani, Lazarides,
  Moessner, and Sondhi}}]{Khemani2016}
\bibinfo{author}{\bibfnamefont{V.}~\bibnamefont{Khemani}},
  \bibinfo{author}{\bibfnamefont{A.}~\bibnamefont{Lazarides}},
  \bibinfo{author}{\bibfnamefont{R.}~\bibnamefont{Moessner}}, \bibnamefont{and}
  \bibinfo{author}{\bibfnamefont{S.~L.} \bibnamefont{Sondhi}},
  \bibinfo{journal}{Phys. Rev. Lett.} \textbf{\bibinfo{volume}{116}},
  \bibinfo{pages}{250401} (\bibinfo{year}{2016}),
  \urlprefix\url{https://link.aps.org/doi/10.1103/PhysRevLett.116.250401}.

\bibitem[{\citenamefont{Yao et~al.}(2017)\citenamefont{Yao, Potter, Potirniche,
  and Vishwanath}}]{Yao2017}
\bibinfo{author}{\bibfnamefont{N.~Y.} \bibnamefont{Yao}},
  \bibinfo{author}{\bibfnamefont{A.~C.} \bibnamefont{Potter}},
  \bibinfo{author}{\bibfnamefont{I.-D.} \bibnamefont{Potirniche}},
  \bibnamefont{and}
  \bibinfo{author}{\bibfnamefont{A.}~\bibnamefont{Vishwanath}},
  \bibinfo{journal}{Phys. Rev. Lett.} \textbf{\bibinfo{volume}{118}},
  \bibinfo{pages}{030401} (\bibinfo{year}{2017}),
  \urlprefix\url{https://link.aps.org/doi/10.1103/PhysRevLett.118.030401}.

\bibitem[{\citenamefont{Wilczek}(2012)}]{Wilczek2012}
\bibinfo{author}{\bibfnamefont{F.}~\bibnamefont{Wilczek}},
  \bibinfo{journal}{Phys. Rev. Lett.} \textbf{\bibinfo{volume}{109}},
  \bibinfo{pages}{160401} (\bibinfo{year}{2012}),
  \urlprefix\url{https://link.aps.org/doi/10.1103/PhysRevLett.109.160401}.

\bibitem[{\citenamefont{Choi et~al.}(2017)\citenamefont{Choi, Choi, Landig,
  Kucsko, Zhou, Isoya, Jelezko, Onoda, Sumiya, Khemani et~al.}}]{Choi2017}
\bibinfo{author}{\bibfnamefont{S.}~\bibnamefont{Choi}},
  \bibinfo{author}{\bibfnamefont{J.}~\bibnamefont{Choi}},
  \bibinfo{author}{\bibfnamefont{R.}~\bibnamefont{Landig}},
  \bibinfo{author}{\bibfnamefont{G.}~\bibnamefont{Kucsko}},
  \bibinfo{author}{\bibfnamefont{H.}~\bibnamefont{Zhou}},
  \bibinfo{author}{\bibfnamefont{J.}~\bibnamefont{Isoya}},
  \bibinfo{author}{\bibfnamefont{F.}~\bibnamefont{Jelezko}},
  \bibinfo{author}{\bibfnamefont{S.}~\bibnamefont{Onoda}},
  \bibinfo{author}{\bibfnamefont{H.}~\bibnamefont{Sumiya}},
  \bibinfo{author}{\bibfnamefont{V.}~\bibnamefont{Khemani}},
  \bibnamefont{et~al.}, \bibinfo{journal}{Nature}
  \textbf{\bibinfo{volume}{543}}, \bibinfo{pages}{221 EP }
  (\bibinfo{year}{2017}),
  \urlprefix\url{http://dx.doi.org/10.1038/nature21426}.

\bibitem[{\citenamefont{Zhang et~al.}(2017)\citenamefont{Zhang, Hess,
  Kyprianidis, Becker, Lee, Smith, Pagano, Potirniche, Potter, Vishwanath
  et~al.}}]{Zhang2017}
\bibinfo{author}{\bibfnamefont{J.}~\bibnamefont{Zhang}},
  \bibinfo{author}{\bibfnamefont{P.~W.} \bibnamefont{Hess}},
  \bibinfo{author}{\bibfnamefont{A.}~\bibnamefont{Kyprianidis}},
  \bibinfo{author}{\bibfnamefont{P.}~\bibnamefont{Becker}},
  \bibinfo{author}{\bibfnamefont{A.}~\bibnamefont{Lee}},
  \bibinfo{author}{\bibfnamefont{J.}~\bibnamefont{Smith}},
  \bibinfo{author}{\bibfnamefont{G.}~\bibnamefont{Pagano}},
  \bibinfo{author}{\bibfnamefont{I.-D.} \bibnamefont{Potirniche}},
  \bibinfo{author}{\bibfnamefont{A.~C.} \bibnamefont{Potter}},
  \bibinfo{author}{\bibfnamefont{A.}~\bibnamefont{Vishwanath}},
  \bibnamefont{et~al.}, \bibinfo{journal}{Nature}
  \textbf{\bibinfo{volume}{543}}, \bibinfo{pages}{217 EP }
  (\bibinfo{year}{2017}),
  \urlprefix\url{http://dx.doi.org/10.1038/nature21413}.

\bibitem[{\citenamefont{von Keyserlingk et~al.}(2016)\citenamefont{von
  Keyserlingk, Khemani, and Sondhi}}]{vonKeyserlingk2016}
\bibinfo{author}{\bibfnamefont{C.~W.} \bibnamefont{von Keyserlingk}},
  \bibinfo{author}{\bibfnamefont{V.}~\bibnamefont{Khemani}}, \bibnamefont{and}
  \bibinfo{author}{\bibfnamefont{S.~L.} \bibnamefont{Sondhi}},
  \bibinfo{journal}{Phys. Rev. B} \textbf{\bibinfo{volume}{94}},
  \bibinfo{pages}{085112} (\bibinfo{year}{2016}),
  \urlprefix\url{https://link.aps.org/doi/10.1103/PhysRevB.94.085112}.

\bibitem[{\citenamefont{Ponte et~al.}(2015)\citenamefont{Ponte,
  Papi\ifmmode~\acute{c}\else \'{c}\fi{}, Huveneers, and Abanin}}]{Ponte2015}
\bibinfo{author}{\bibfnamefont{P.}~\bibnamefont{Ponte}},
  \bibinfo{author}{\bibfnamefont{Z.}~\bibnamefont{Papi\ifmmode~\acute{c}\else
  \'{c}\fi{}}}, \bibinfo{author}{\bibfnamefont{F.~m.~c.}
  \bibnamefont{Huveneers}}, \bibnamefont{and}
  \bibinfo{author}{\bibfnamefont{D.~A.} \bibnamefont{Abanin}},
  \bibinfo{journal}{Phys. Rev. Lett.} \textbf{\bibinfo{volume}{114}},
  \bibinfo{pages}{140401} (\bibinfo{year}{2015}),
  \urlprefix\url{https://link.aps.org/doi/10.1103/PhysRevLett.114.140401}.

\bibitem[{\citenamefont{Lazarides et~al.}(2015)\citenamefont{Lazarides, Das,
  and Moessner}}]{Lazarides2015}
\bibinfo{author}{\bibfnamefont{A.}~\bibnamefont{Lazarides}},
  \bibinfo{author}{\bibfnamefont{A.}~\bibnamefont{Das}}, \bibnamefont{and}
  \bibinfo{author}{\bibfnamefont{R.}~\bibnamefont{Moessner}},
  \bibinfo{journal}{Phys. Rev. Lett.} \textbf{\bibinfo{volume}{115}},
  \bibinfo{pages}{030402} (\bibinfo{year}{2015}),
  \urlprefix\url{https://link.aps.org/doi/10.1103/PhysRevLett.115.030402}.

\bibitem[{\citenamefont{Bordia et~al.}(2017)\citenamefont{Bordia, L{\"u}schen,
  Schneider, Knap, and Bloch}}]{Bordia2017}
\bibinfo{author}{\bibfnamefont{P.}~\bibnamefont{Bordia}},
  \bibinfo{author}{\bibfnamefont{H.}~\bibnamefont{L{\"u}schen}},
  \bibinfo{author}{\bibfnamefont{U.}~\bibnamefont{Schneider}},
  \bibinfo{author}{\bibfnamefont{M.}~\bibnamefont{Knap}}, \bibnamefont{and}
  \bibinfo{author}{\bibfnamefont{I.}~\bibnamefont{Bloch}},
  \bibinfo{journal}{Nature Physics} \textbf{\bibinfo{volume}{13}},
  \bibinfo{pages}{460 EP } (\bibinfo{year}{2017}), \bibinfo{note}{article},
  \urlprefix\url{http://dx.doi.org/10.1038/nphys4020}.

\bibitem[{\citenamefont{Nielsen}(2015)}]{Nielsen2015}
\bibinfo{author}{\bibfnamefont{M.}~\bibnamefont{Nielsen}},
  \emph{\bibinfo{title}{{Neural Networks and Deep Learning}}}
  (\bibinfo{publisher}{Determination Press}, \bibinfo{year}{2015}).

\bibitem[{\citenamefont{Hochreiter and
  Schmidhuber}(1997)}]{HochreiterSchmidhuber}
\bibinfo{author}{\bibfnamefont{S.}~\bibnamefont{Hochreiter}} \bibnamefont{and}
  \bibinfo{author}{\bibfnamefont{J.}~\bibnamefont{Schmidhuber}},
  \bibinfo{journal}{Neural computation} \textbf{\bibinfo{volume}{9}},
  \bibinfo{pages}{1735} (\bibinfo{year}{1997}).

\bibitem[{\citenamefont{{Guo} et~al.}(2017)\citenamefont{{Guo}, {Pleiss},
  {Sun}, and {Weinberger}}}]{Calibration}
\bibinfo{author}{\bibfnamefont{C.}~\bibnamefont{{Guo}}},
  \bibinfo{author}{\bibfnamefont{G.}~\bibnamefont{{Pleiss}}},
  \bibinfo{author}{\bibfnamefont{Y.}~\bibnamefont{{Sun}}}, \bibnamefont{and}
  \bibinfo{author}{\bibfnamefont{K.~Q.} \bibnamefont{{Weinberger}}},
  \bibinfo{journal}{ArXiv e-prints}  (\bibinfo{year}{2017}),
  \eprint{1706.04599}.

\bibitem[{\citenamefont{Srivastava et~al.}(2014)\citenamefont{Srivastava,
  Hinton, Krizhevsky, Sutskever, and Salakhutdinov}}]{Srivastava2014}
\bibinfo{author}{\bibfnamefont{N.}~\bibnamefont{Srivastava}},
  \bibinfo{author}{\bibfnamefont{G.~E.} \bibnamefont{Hinton}},
  \bibinfo{author}{\bibfnamefont{A.}~\bibnamefont{Krizhevsky}},
  \bibinfo{author}{\bibfnamefont{I.}~\bibnamefont{Sutskever}},
  \bibnamefont{and}
  \bibinfo{author}{\bibfnamefont{R.}~\bibnamefont{Salakhutdinov}},
  \bibinfo{journal}{Journal of machine learning research}
  \textbf{\bibinfo{volume}{15}}, \bibinfo{pages}{1929} (\bibinfo{year}{2014}).

\bibitem[{\citenamefont{Kingma and Ba}(2015)}]{Kingma2015}
\bibinfo{author}{\bibfnamefont{D.~P.} \bibnamefont{Kingma}} \bibnamefont{and}
  \bibinfo{author}{\bibfnamefont{J.~L.} \bibnamefont{Ba}}, pp.
  \bibinfo{pages}{1--15} (\bibinfo{year}{2015}), ISSN \bibinfo{issn}{09252312},
  \eprint{1412.6980v9}, \urlprefix\url{http://arxiv.org/abs/1412.6980}.

\bibitem[{\citenamefont{Pal and Huse}(2010)}]{Pal2010}
\bibinfo{author}{\bibfnamefont{A.}~\bibnamefont{Pal}} \bibnamefont{and}
  \bibinfo{author}{\bibfnamefont{D.~A.} \bibnamefont{Huse}},
  \bibinfo{journal}{Phys. Rev. B} \textbf{\bibinfo{volume}{82}},
  \bibinfo{pages}{174411} (\bibinfo{year}{2010}).

\bibitem[{\citenamefont{Luitz et~al.}(2015)\citenamefont{Luitz, Laflorencie,
  and Alet}}]{Luitz2015}
\bibinfo{author}{\bibfnamefont{D.~J.} \bibnamefont{Luitz}},
  \bibinfo{author}{\bibfnamefont{N.}~\bibnamefont{Laflorencie}},
  \bibnamefont{and} \bibinfo{author}{\bibfnamefont{F.}~\bibnamefont{Alet}},
  \bibinfo{journal}{Physical Review B} \textbf{\bibinfo{volume}{91}},
  \bibinfo{pages}{081103} (\bibinfo{year}{2015}),
  \urlprefix\url{https://journals.aps.org/prb/abstract/10.1103/PhysRevB.91.081103}.

\bibitem[{\citenamefont{Serbyn et~al.}(2015)\citenamefont{Serbyn, Papi{\'{c}},
  and Abanin}}]{Serbyn2015}
\bibinfo{author}{\bibfnamefont{M.}~\bibnamefont{Serbyn}},
  \bibinfo{author}{\bibfnamefont{Z.}~\bibnamefont{Papi{\'{c}}}},
  \bibnamefont{and} \bibinfo{author}{\bibfnamefont{D.~A.}
  \bibnamefont{Abanin}}, \bibinfo{journal}{Phys. Rev. X}
  \textbf{\bibinfo{volume}{5}}, \bibinfo{pages}{41047} (\bibinfo{year}{2015}).

\bibitem[{\citenamefont{Bera et~al.}(2015)\citenamefont{Bera, Schomerus,
  Heidrich-Meisner, and Bardarson}}]{Bera2015}
\bibinfo{author}{\bibfnamefont{S.}~\bibnamefont{Bera}},
  \bibinfo{author}{\bibfnamefont{H.}~\bibnamefont{Schomerus}},
  \bibinfo{author}{\bibfnamefont{F.}~\bibnamefont{Heidrich-Meisner}},
  \bibnamefont{and} \bibinfo{author}{\bibfnamefont{J.~H.}
  \bibnamefont{Bardarson}}, \bibinfo{journal}{Phys. Rev. Lett.}
  \textbf{\bibinfo{volume}{115}}, \bibinfo{pages}{046603}
  (\bibinfo{year}{2015}),
  \urlprefix\url{https://link.aps.org/doi/10.1103/PhysRevLett.115.046603}.

\bibitem[{\citenamefont{Iyer et~al.}(2013)\citenamefont{Iyer, Oganesyan,
  Refael, and Huse}}]{Iyer2013}
\bibinfo{author}{\bibfnamefont{S.}~\bibnamefont{Iyer}},
  \bibinfo{author}{\bibfnamefont{V.}~\bibnamefont{Oganesyan}},
  \bibinfo{author}{\bibfnamefont{G.}~\bibnamefont{Refael}}, \bibnamefont{and}
  \bibinfo{author}{\bibfnamefont{D.~A.} \bibnamefont{Huse}},
  \bibinfo{journal}{Phys. Rev. B} \textbf{\bibinfo{volume}{87}},
  \bibinfo{pages}{134202} (\bibinfo{year}{2013}),
  \urlprefix\url{https://link.aps.org/doi/10.1103/PhysRevB.87.134202}.

\bibitem[{\citenamefont{Gopalakrishnan
  et~al.}(2016)\citenamefont{Gopalakrishnan, Knap, and
  Demler}}]{Gopalakrishnan2016}
\bibinfo{author}{\bibfnamefont{S.}~\bibnamefont{Gopalakrishnan}},
  \bibinfo{author}{\bibfnamefont{M.}~\bibnamefont{Knap}}, \bibnamefont{and}
  \bibinfo{author}{\bibfnamefont{E.}~\bibnamefont{Demler}},
  \bibinfo{journal}{Phys. Rev. B} \textbf{\bibinfo{volume}{94}},
  \bibinfo{pages}{094201} (\bibinfo{year}{2016}),
  \urlprefix\url{https://link.aps.org/doi/10.1103/PhysRevB.94.094201}.

\bibitem[{\citenamefont{Schreiber et~al.}(2015)\citenamefont{Schreiber,
  Hodgman, Bordia, Luschen, Fischer, Vosk, Altman, Schneider, and
  Bloch}}]{Schreiber2015}
\bibinfo{author}{\bibfnamefont{M.}~\bibnamefont{Schreiber}},
  \bibinfo{author}{\bibfnamefont{S.~S.} \bibnamefont{Hodgman}},
  \bibinfo{author}{\bibfnamefont{P.}~\bibnamefont{Bordia}},
  \bibinfo{author}{\bibfnamefont{H.~P.} \bibnamefont{Luschen}},
  \bibinfo{author}{\bibfnamefont{M.~H.} \bibnamefont{Fischer}},
  \bibinfo{author}{\bibfnamefont{R.}~\bibnamefont{Vosk}},
  \bibinfo{author}{\bibfnamefont{E.}~\bibnamefont{Altman}},
  \bibinfo{author}{\bibfnamefont{U.}~\bibnamefont{Schneider}},
  \bibnamefont{and} \bibinfo{author}{\bibfnamefont{I.}~\bibnamefont{Bloch}},
  \bibinfo{journal}{Science} \textbf{\bibinfo{volume}{349}},
  \bibinfo{pages}{842} (\bibinfo{year}{2015}), ISSN \bibinfo{issn}{0036-8075},
  \eprint{1501.05661v1},
  \urlprefix\url{http://www.sciencemag.org/cgi/doi/10.1126/science.aaa7432}.

\bibitem[{\citenamefont{Luitz et~al.}(2016)\citenamefont{Luitz, Laflorencie,
  and Alet}}]{Luitz2016}
\bibinfo{author}{\bibfnamefont{D.~J.} \bibnamefont{Luitz}},
  \bibinfo{author}{\bibfnamefont{N.}~\bibnamefont{Laflorencie}},
  \bibnamefont{and} \bibinfo{author}{\bibfnamefont{F.}~\bibnamefont{Alet}},
  \bibinfo{journal}{Phys. Rev. B} \textbf{\bibinfo{volume}{93}},
  \bibinfo{pages}{060201} (\bibinfo{year}{2016}),
  \urlprefix\url{https://link.aps.org/doi/10.1103/PhysRevB.93.060201}.

\bibitem[{\citenamefont{{Gupta} and {Biercuk}}(2017)}]{Gupta2017}
\bibinfo{author}{\bibfnamefont{R.}~\bibnamefont{{Gupta}}} \bibnamefont{and}
  \bibinfo{author}{\bibfnamefont{M.~J.} \bibnamefont{{Biercuk}}},
  \bibinfo{journal}{ArXiv e-prints}  (\bibinfo{year}{2017}),
  \eprint{1712.01291}.

\bibitem[{\citenamefont{Abadi et~al.}(2016)}]{Abadi2016}
\bibinfo{author}{\bibfnamefont{M.}~\bibnamefont{Abadi}} \bibnamefont{et~al.},
  \bibinfo{journal}{arXiv:1603.04467}  (\bibinfo{year}{2016}),
  \eprint{1603.04467}, \urlprefix\url{http://arxiv.org/abs/1603.04467}.

\bibitem[{\citenamefont{Chollet et~al.}(2015)}]{chollet2015keras}
\bibinfo{author}{\bibfnamefont{F.}~\bibnamefont{Chollet}} \bibnamefont{et~al.},
  \emph{\bibinfo{title}{Keras}},
  \bibinfo{howpublished}{\url{https://github.com/fchollet/keras}}
  (\bibinfo{year}{2015}).

\bibitem [{\citenamefont {{Guo}}\ \emph {et~al.}(2017)\citenamefont {{Guo}},
  \citenamefont {{Pleiss}}, \citenamefont {{Sun}},\ and\ \citenamefont
  {{Weinberger}}}]{Calibration}%
  \BibitemOpen
  \bibfield  {author} {\bibinfo {author} {\bibfnamefont {C.}~\bibnamefont
  {{Guo}}}, \bibinfo {author} {\bibfnamefont {G.}~\bibnamefont {{Pleiss}}},
  \bibinfo {author} {\bibfnamefont {Y.}~\bibnamefont {{Sun}}}, \ and\ \bibinfo
  {author} {\bibfnamefont {K.~Q.}\ \bibnamefont {{Weinberger}}},\ }\href@noop
  {http://arxiv.org/abs/1706.04599} {\bibfield  {journal} {\bibinfo  {journal} {ArXiv e-prints}\ } (\bibinfo
  {year} {2017})}

\bibitem [{\citenamefont {{Zhang}}\ \emph {et~al.}(2017)\citenamefont {{Zhang}},
  \citenamefont {{Bengio}}, \citenamefont {{Hardt}}, \citenamefont {{Recht}},\ and\ \citenamefont
  {{Vinyals}}}]{Generalization}%
  \BibitemOpen
  \bibfield  {author} {\bibinfo {author} {\bibfnamefont {C.}~\bibnamefont
  {{Zhang}}}, \bibinfo {author} {\bibfnamefont {S.}~\bibnamefont {{Bengio}}}, \bibinfo {author} {\bibfnamefont {M.}~\bibnamefont {{Hardt}}} \bibinfo {author} {\bibfnamefont {B.}~\bibnamefont {{Recht}}}, \ and\ \bibinfo
  {author} {\bibfnamefont {O.}\ \bibnamefont {{Vinyals}}},\ }\href@noop
  http://arxiv.org/abs/1611.03530{} {\bibfield  {journal} {\bibinfo  {journal} {ArXiv e-prints}\ } (\bibinfo
  {year} {2016})}

\end{thebibliography}

\vspace{-0.5cm}
\section*{Acknowledgments} 
\vspace{-0.2cm} E.v.N. gratefully acknowledges financial support from the Swiss National Science Foundation through grant P2EZP2-172185. E.v.N. also acknowledges fruitful discussions with Manuel Endres. E.B. is grateful to Netanel Lindner for his support and acknowledges financial support from the European Research Council (ERC) under the European Union Horizon  2020  Research  and  Innovation  Programme  (GrantAgreement No.  639172). G.R. is grateful to the the NSF for funding through the grant DMR-1040435 as well as the Packard Foundation. We are grateful for support
from the IQIM, an NSF physics frontier center funded in part by the Moore Foundation. The authors used the TensorFlow \cite{Abadi2016} backend for Keras~\cite{chollet2015keras}. E.v.N. and E.B. contributed equally to this work.

\clearpage
\newpage

\onecolumngrid
\vspace*{\stretch{1.0}}
\begin{center}
\Large\textbf{Supplementary Material}\\
\vspace{0.5cm}
\end{center}
\vspace*{\stretch{2.0}}

\twocolumngrid

\section*{Training LSTMs on a single spin}
To gain a better understanding of what the LSTM neurons are extracting from the magnetization input, we focus here on the case of a single spin.
In this section, we consider the magnetization $m_i(t)$ of one spin close to the center of our $20$ site chain, i.e. $i = 10$. We restrict ourselves to initial states
for which the energy density $\varepsilon = 0.5$.

\begin{figure}[!h]
	\includegraphics[width=1\columnwidth]{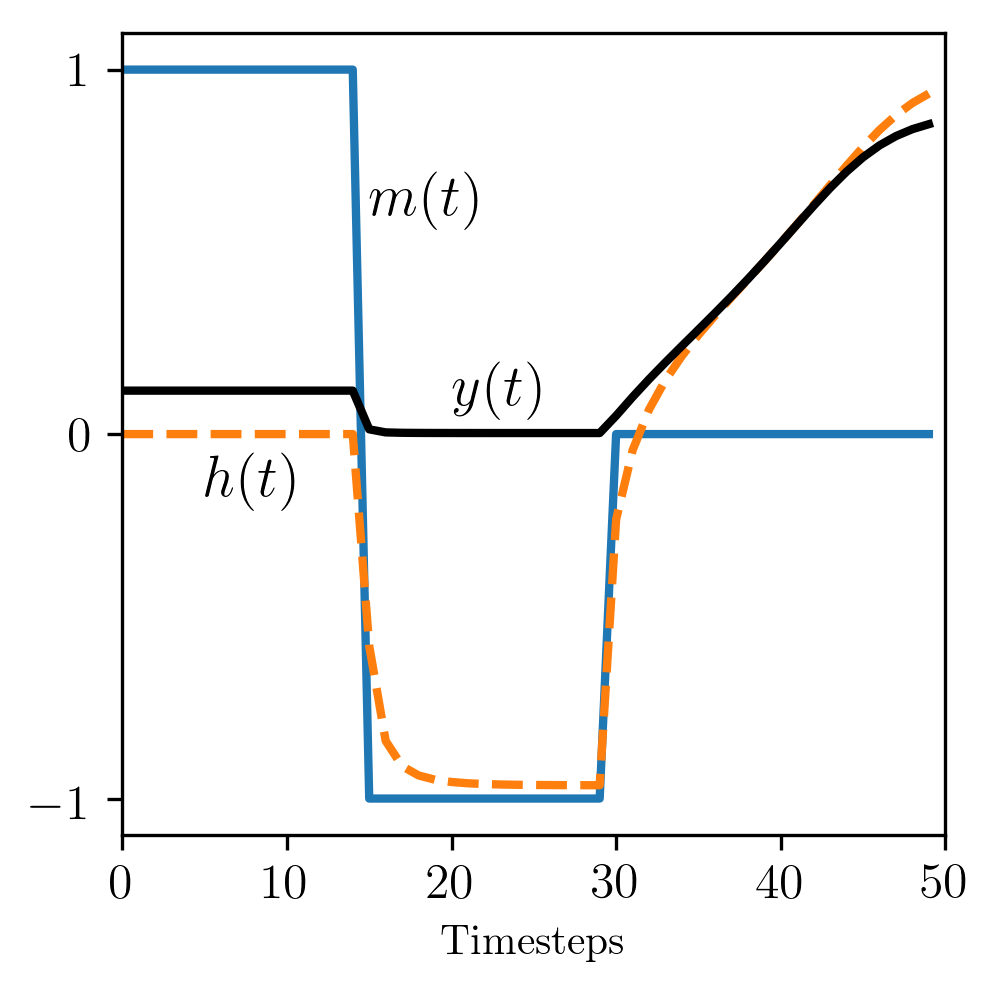}
	\caption{A single handcrafted LSTM neuron, modeled after the observed behaviour of a trained LSTM neuron. To clearly demonstrate the various regimes of the neuron, we evaluate its hidden state $h(t)$ and the resulting prediction $y(t)$ (where $y(t) \approx 0$ corresponds to MBL and $y(t) \approx 1$ corresponds to thermalizing behaviour) for a specific magnetization signal $m(t)$. Whenever the magnetization is close to $0$, the neuron builds confidence that the signal classifies as a thermalizing signal, whereas a magnetization that sticks around extreme values signals localization.} \label{fig:singlespinLSTM}
\end{figure}

We start by training a single LSTM unit on this single spin data using the blanking setup described in the main text. After training, we analyzed the hidden state $h(t)$ and the LSTM output $y(t)$ on various magnetization curves. It is particularly instructive however to evaluate the neuron's behaviour on a handcrafted magnetization signal. Even though re-training this neuron (starting with random initial internal weights) results in different quantitative behaviour, it seems to always show qualitatively similar response. That is, the neuron output effectively functions as an integrator depending on the magnetization, as exemplified in Fig.~\ref{fig:singlespinLSTM}. Intuitively, if the magnetization sticks to zero, the neuron slowly becomes more certain of ETH behaviour. However, having a magnetization that is constantly close to $\pm 1$, builds confidence for the MBL classification. Based on these observations, the behaviour shown in Fig.~\ref{fig:singlespinLSTM} is for a handcrafted LSTM neuron designed to implement this integration behavior.

Training multiple LSTM units on a single spin is also beneficial for understanding, and shows that different LSTM units all essentially behave as similar integrators, but they integrate different parts of the signal. Combining the results of all these neurons together leads to a more robust prediction for the transition point w.r.t. $W$, as judged from the scaling analysis of the number of LSTM units (see main text).

\section*{Robustness of blanking}
In this section, we discuss several approaches to assess the validity of the network's predictions in the untrained regions of the phase space.

\subsection{Blanking region dependence}
\begin{figure}[!h]
	\includegraphics[width=1\columnwidth]{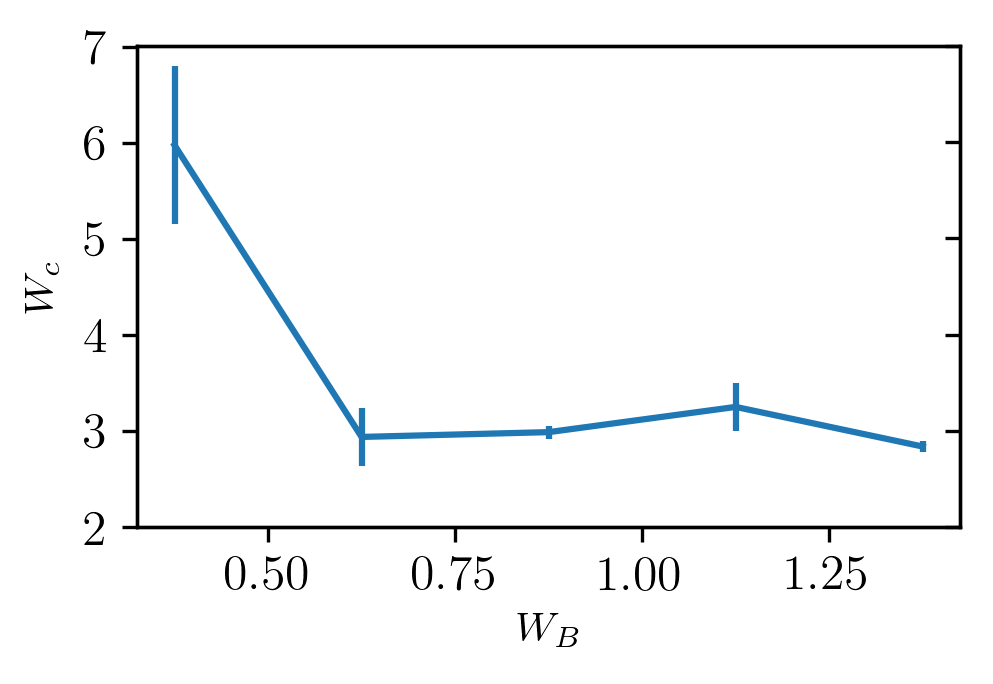}
	\caption{The predicted $W_c$ after 25 epochs of training, averaged over 10 re-trained networks with the same parameters as in the main text, as extracted from the peaks in the confusion $C$ (see main text). The x-axis shows $W_B$ representing the regions of the phase space, namely $[W_B, 8-W_B]$, that were used to train the network. Importantly, from $W_B \gtrsim 0.75$ the predicted $W_c$ shows little variation and reasonably small variance, indicating that the extracted phase boundary is insensitive to the amount of data included (symmetrically) from both ends of the phase space. } \label{fig:finitedatascaling}
\end{figure}

First, we examine the sensitivity of the predicted phase transition to the size of the region in phase space where the network is trained on. On the one hand, one would like the network to rely on as little human input as possible; on the other hand, a network trained on very narrow regions in phase space might pick up features that are specific to that region rather than universal to the phase that region belongs to. We therefore repeat our training process for different blanked-out regions of the form $[W_B, 8-W_B]$. For each blanked out region we train 10 networks and extract the $W_c$ (from the peak in the confusion $C$) predicted by each. Figure~\ref{fig:finitedatascaling} shows the mean $W_c$ as well as the standard deviation as a function of the length of the training region $W_B$. We observe that the predictions for $W_c$ converge to a narrow range when $W_B \gtrsim 0.75$ and is weakly dependent on $W_B$ for larger values.

We wish to remark here finally, that both the $l_2$ regularization and the confidence enhancement affect the shape of the learned MBL phase diagram. In their absence, we observe a weaker dependence of the critical disorder strength on the energy density.

\subsection*{Calibration}
\begin{figure}[h]
	\centering
	\includegraphics[width=1\columnwidth]{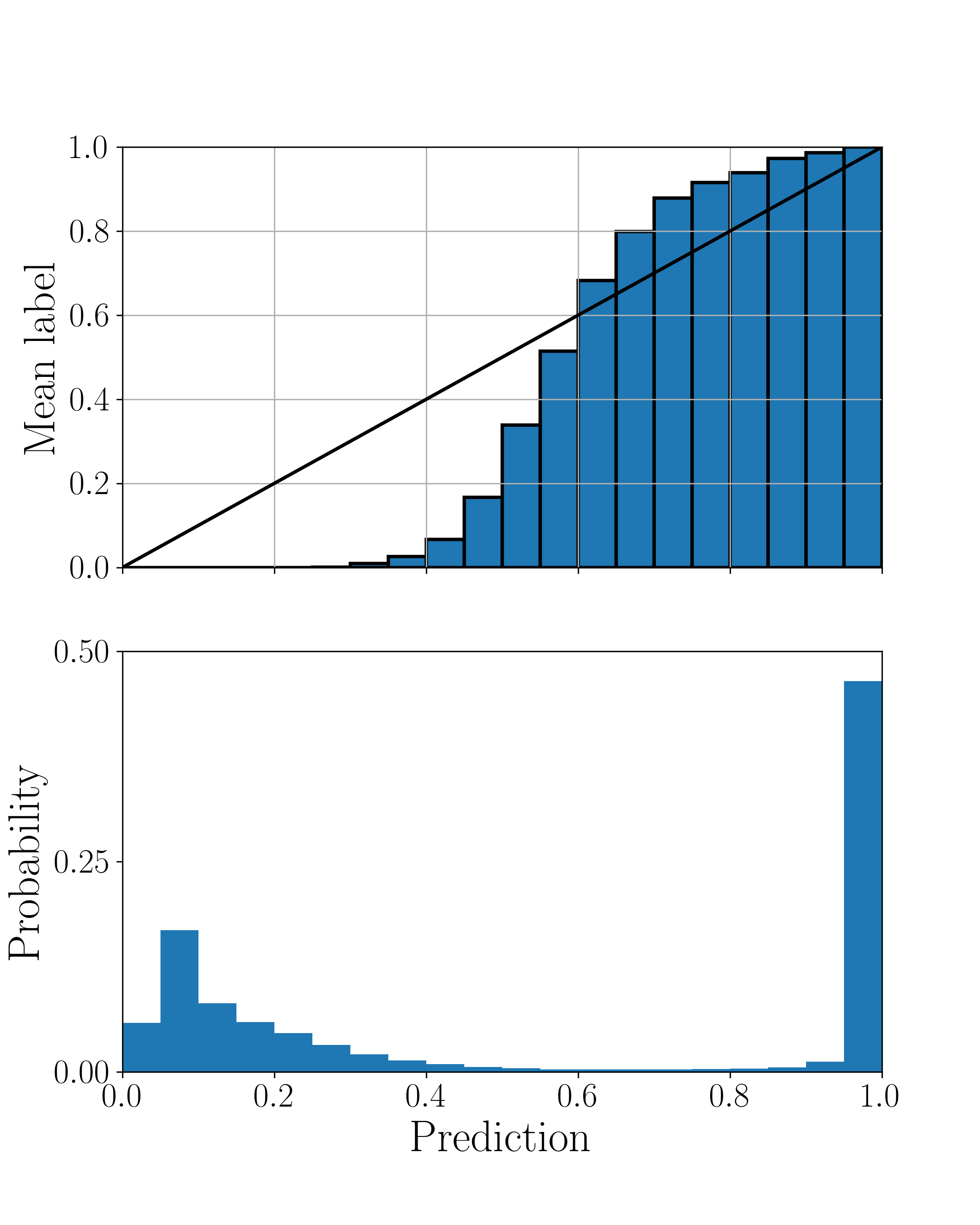}
	\caption{A closer look at network performance in a region where the correct phase is known. We train 100 networks on the regions $W\in[0,0.5]$ and $W\in[7.5,8]$ in the random-field Heisenberg model at energy density $\epsilon=0.5$, and evaluate their performance on additional samples from these parameter ranges as well as the $W\in[0,0.5]$ and $W\in[7.5,8]$. Top panel shows the mean label (0 corresponds to MBL, 1 to ergodic) as a function of network prediction, averaged over the all the predictions of the different networks in equal intervals of length 0.05. Bottom panel shows the probability for each prediction to occur in each of these intervals. We observe a higher mean confidence for ergodic samples in this regime compared to the MBL samples.
	\label{fig:Calibration} }
\end{figure}

We also examine the network's calibration, namely how well its confidence corresponds to its accuracy \cite{Calibration}. 
We train a network on the random-field Heisenberg model at energy density $\epsilon=0.5$ in the regions $W\in[0,0.5]$ and $W\in[7.5,8]$ (as in the main text), and then analyze its predictions on a calibration set containing the test set (different disorder realizations taken from the training region) as well as the untrained regions $W\in[0.5,1]$ and $W\in[7,7.5]$. To collect a smooth histogram, we repeat this process over 100 networks. Fig. ~\ref{fig:Calibration} shows the probability that an instance belongs to the ergodic phase as a function of the output of its corresponding network (top panel), as well as a histogram of the outputs of the networks (bottom panel). We observe that in these regions of phase space the networks are mostly underconfident, assigning a lower confidence value than their actual accuracy. This is especially true for the MBL phase, where the networks are on average less confident than for the ergodic phase.

\subsection*{Unseen phases}
\begin{figure}[h]
	\includegraphics[width=1\columnwidth]{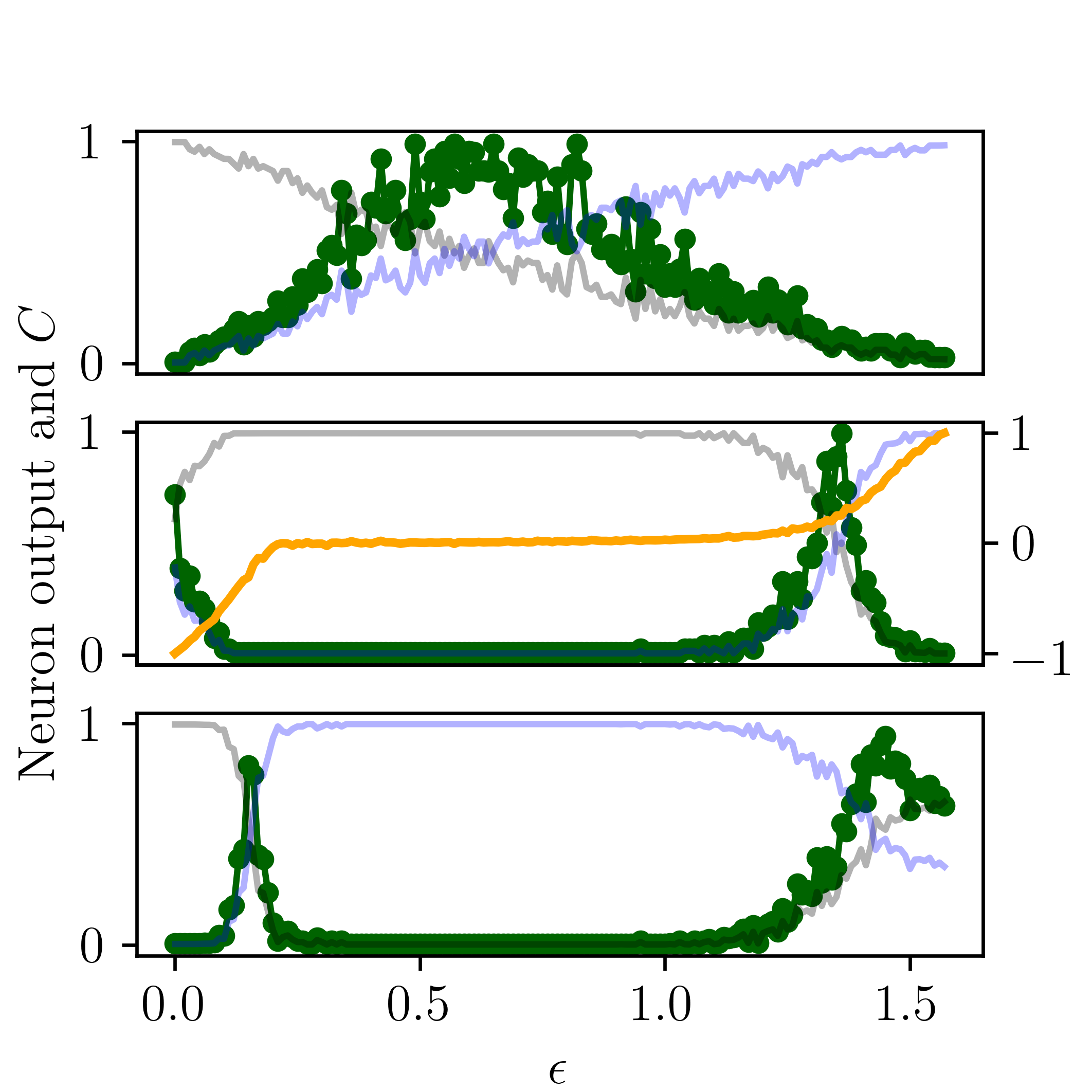}
	\caption{Networks assign a confused output to phases they had not encountered during training. We train networks to distinguish between two of the three phases of the model featured in Fig. 3 in the main text. In the top panel, a network is trained on the two extremes of the parameter range (time-crystal at $\epsilon=0$ and Floquet-MBL at $\epsilon=\pi$). When evaluated, the network does not assign a confident prediction to any of these two phases throughout most of the parameter range, indicating the existance of a third, unseen phase - the ergodic phase. Similarly, in the middle panel a network is trained only on the ergodic and Floquet-MBL phases, and in the bottom panel on the time-crystal and ergodic phases; each network is baffled by the phase it was not trained on. The middle panel also shows the averaged final imbalance throughout the parameter range (yellow line) as a rough indicator of the correct phase.}	\label{fig:Unseen_phase} 
\end{figure}

Finally, it is interesting to examine how the network reacts to a phase it had not encountered during training. Focusing now on our time-dependent model, we train a network for the binary classification task of distinguishing between the Floquet-MBL and the time-crystal phases, without exposing it to any data from the ergodic phase during training. Rather than confidently misclassifying this unseen phase, the network assigns a highly confused output to most of the ergodic phase (Fig. ~\ref{fig:Unseen_phase}). This suggests that the network is robust to adversarial examples; moreover, it suggests that our method can be used not only for detecting phase transitions between known phases, but also in a semi-unsupervised matter for finding new phases.

\vspace{-0.4cm}


\section*{Comparison with Imbalance}
\begin{figure}[h]
	\includegraphics[width=0.8\textwidth]{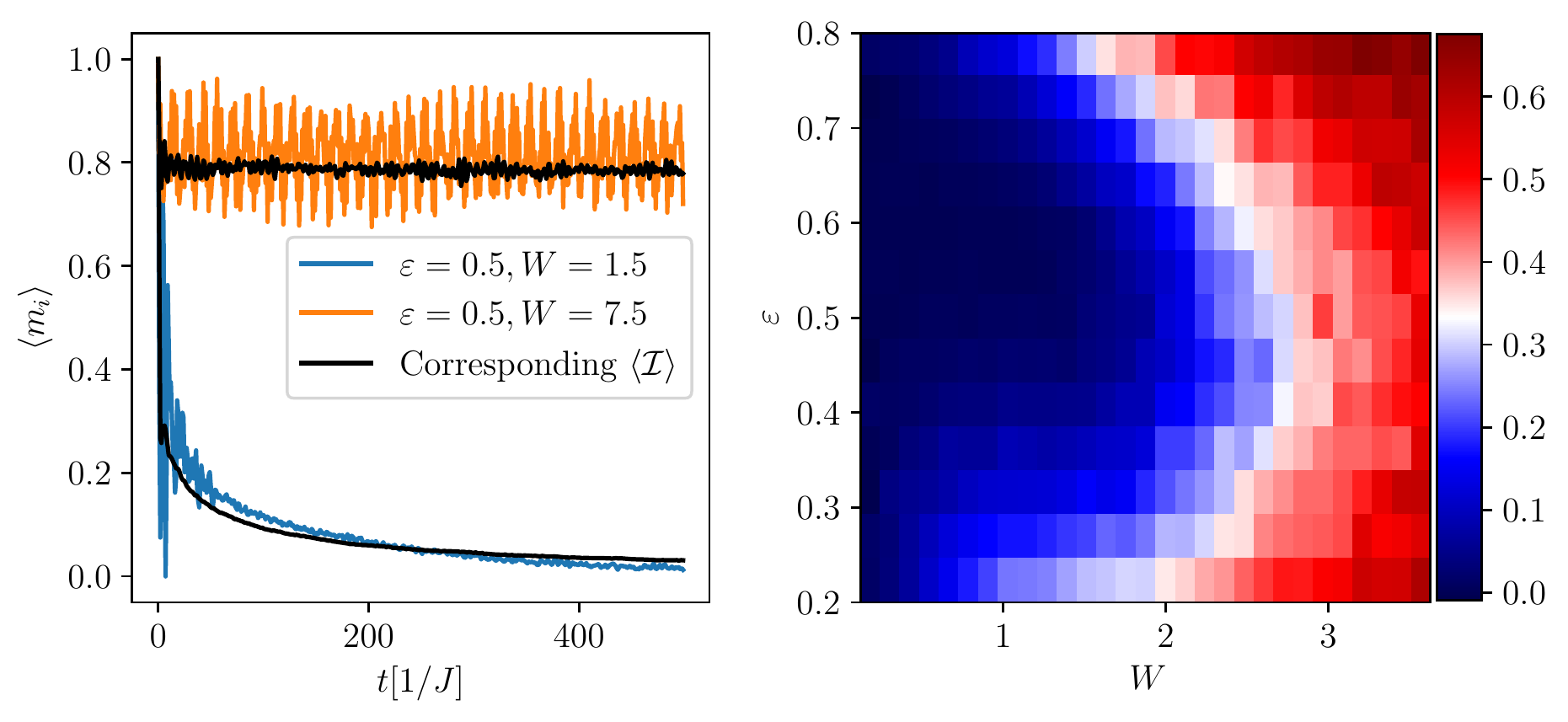}
	\caption{The left panel shows time traces of the magnetization of a central spin ($i=8$) in a $20$-site spin chain, for a given disorder realization and parameters indicated. For weak disorder ($W=1.5$) the spin quickly relaxes, whereas for strong disorder it mostly retains its initial value. The disorder averaged imbalance $\mathcal{I}(t)$ for the corresponding parameter regimes is shown superimposed, indicating that these behaviours are typical for the spins and given regimes. The right panel shows the imbalance at the final time $t_{\textrm{final}} = 500$ as a function of energy density and disorder strength. The resulting phase diagram is suggestive, but putting a phase boundary via a threshold on $\mathcal{I}$ is ambiguous.
	 }\label{fig:spinandimbalance}
\end{figure}

To illustrate the ambiguity in using the imbalance $\mathcal{I}$ as a transition indicator, consider the MBL system discussed in the main text.
In the left panel of Fig.~\ref{fig:spinandimbalance} we show the traces of a single spin in the middle of the sample (for a single disorder realization), at energy density $\varepsilon = 0.5$ and at two different disorder strengths. For weak disorder, the spin quickly relaxes whereas for strong disorder the spin maintains a value close to its initial one for the entire simulated time $t_{\textrm{final}}$. Superimposed on top of these example curves is the disorder averaged imbalance $\langle \mathcal{I}(t) \rangle$.

Clearly, the disorder averaged imbalance at the final simulation time $\langle \mathcal{I}(t_{\textrm{final}})\rangle$ easily distinguishes between the extreme regimes of weak and strong disorder. We know that in those regimes the model is ergodic and non-ergodic, respectively, so we might naively use this quantity to compute a phase diagram as in the rightmost panel of Fig.~\ref{fig:spinandimbalance}. Although suggestive, a criterion for the phase boundary from $\mathcal{I}(t)$ (for $t = 500$ in this case) is non-trivial. If we had access to infinite times, the remaining imbalance would be a clear criterion; but for finite time data as in an experiment, one would need to extrapolate. What is more, is that a finite size scaling attempt on the imbalance does not show a crossing point~\cite{Iyer2013}. Rather, here we suggest a method that, given the physics of the extreme regimes, consistently finds such a threshold between the strong and weak disorder regimes from the data only.

\section*{Training procedure}
For the detection of the MBL transition in the main text, the training was performed as mentioned for $W\in{0.125,0.25,0.375,0.5,7.625,7.5,7.875,8}$. At $\epsilon=0.5$, we generated 100 realizations for each disorder strength, 60 for training and 40 for test. Each realization randomized both disorder and initial spin configuration. The training and test sets therefore consisted of 480 and 320 samples respectively, though effectively the 5600 samples at $0.625 \leq W \leq 7.5$ are utilized during training due to the confidence enhancement~\cite{Schindler2017}. Note that within the blanking framework we can only include samples from the labeled regions in the test set, since we cannot assess the network's accuracy in regions where the correct label is unknown. The assessment of overfitting is therefore limited to the training region, and needs to be interpreted slightly differently from the usual case~\cite{Generalization}. Namely, the generalization we wish to test is that of whether the network managed to extract the right physical model from the training regions only. A perfect fit of the data is hence not necessarily a bad thing, if indeed one can show that the data in these regions is in principle enough to extract the correct model. Such statements might be testable using adversarial perturbations, but we have not investigated this direction. Regardless, within the training regions we observed that the training-set and test-set accuracies increased with the number of LSTM units in the range we examined. For $2$, $4$, $16$ and $32$ units respectively, averaged over 10 retrainings, we obtained final training accuracies $0.94$, $0.99$, $1$, $1$ and corresponding test-set accuracies $0.89$, $0.97$, $0.99$, $0.99$.

\onecolumngrid
\noindent%

\end{document}